\documentclass[a4paper,12pt]{article}
\usepackage{jheppub}

\usepackage{epsfig}
\usepackage{amssymb,amsmath}
\usepackage{pstricks}
\usepackage{pst-all}
\usepackage[utf8]{inputenc}
\usepackage{epstopdf}
\usepackage{tikz}

\newcommand{\bea}{\begin{eqnarray}}
\newcommand{\eea}{\end{eqnarray}}
\newcommand{\beq}{\begin{equation}}
\newcommand{\eeq}{\end{equation}}
\newcommand{\nn}{\nonumber}

\newcommand{\tr}{\mathrm{Tr}}

\def\lsi{\raise0.3ex\hbox{$<$\kern-0.75em\raise-1.1ex\hbox{$\sim$}}}
\def\gsi{\raise0.3ex\hbox{$>$\kern-0.75em\raise-1.1ex\hbox{$\sim$}}}
\newcommand{\lsim}{\mathop{\lsi}}

\title{Heavy dense QCD and nuclear matter from an effective lattice theory}
\author{Jens Langelage$^1$,}
\author{Mathias Neuman$^2$,}
\author{Owe Philipsen$^2$}
\affiliation{$^1$: Institute for Theoretical Physics, ETH Z\"urich, CH-8093 Z\"urich, Switzerland}
\affiliation{$^2$: Institut f\"ur Theoretische Physik, Goethe-Universit\"at Frankfurt,\\
Max-von-Laue-Str.~1, 60438 Frankfurt am Main, Germany}
\emailAdd{ljens@phys.ethz.ch}
\emailAdd{neuman@th.physik.uni-frankfurt.de}
\emailAdd{philipsen@th.physik.uni-frankfurt.de}
\abstract{
A three-dimensional effective lattice theory of Polyakov loops is derived from QCD by 
expansions in the fundamental character of the gauge action, $u$, and the hopping parameter, $\kappa$, 
whose action is correct to $\kappa^n u^m$ with $n+m=4$. At finite baryon density,
the effective theory has a sign problem which meets all criteria to be simulated by complex Langevin
as well as by Monte Carlo on small volumes. 
The theory is valid for the thermodynamics of heavy quarks, where its predictions agree with 
simulations of full QCD at zero and imaginary chemical potential. In its region of convergence,
it is moreover amenable to perturbative calculations in the small effective couplings.
In this work we study the challenging cold and dense regime. We find unambiguous
evidence for the nuclear liquid gas transition once the baryon chemical potential approaches
the baryon mass, and calculate the nuclear equation of state.
In particular, we find a negative binding energy per nucleon causing the condensation,  
whose absolute value decreases exponentially as mesons get heavier. For decreasing meson mass,
we observe a first order liquid gas transition with an endpoint at some finite temperature, as
well as gap between the onset of isospin and baryon condensation.
}

\begin{document}
\maketitle

\section{Introduction}

The phase diagram of QCD at finite temperature and baryon density is still largely unknown 
today, because
lattice QCD suffers from a severe sign problem when chemical potential for baryon number 
is non-vanishing. 
Several methods have been devised to circumvent this
obstacle (see e.g. \cite{deForcrand:2010ys} and references therein), but all of them introduce
additional approximations that are valid for small quark chemical potentials only, $\mu/T\lesssim1$. 
In order to reach higher chemical potentials and/or low temperatures,
methods are required which at least potentially may solve this 
problem. Among these are Complex Langevin Dynamics (CLD) 
\cite{Aarts:2013bla,Aarts:2013uxa}, 
transformation of the degrees of freedom into so-called dual variables as demonstrated in scalar models
\cite{Gattringer:2012df,Delgado:2012tm},
and the formulation of quantum field theories on a Lefshetz thimble \cite{Cristoforetti:2012su}. 
In particular, CLD has recently been applied to full QCD in a previously inaccessible 
parameter range \cite{denes}.
However, even if 
an approach should finally succeed in solving the sign problem, it will remain very
hard to study the regime of cold and dense matter. This is because, in order to avoid 
the limiting artifact of saturation at finite lattice spacing, very fine
lattices are required for high density, which implies in turn very 
large temporal lattice extents near $T=0$. 

In this work we further elaborate on  
an effective theory approach \cite{Langelage:2010yr,Fromm:2011qi,Fromm:2012eb,procs}, 
where analytic strong coupling and hopping expansion methods are used to derive
an effective lattice action whose numerical simulation is feasible also in the cold and dense regime.
The sign problem can be handled by complex Langevin simulations in a controlled way, and in certain
parameter ranges even Monte Carlo simulations are possible. Moreover, the effective action
resembles a three-dimensional spin model, such that the numerical effort is vastly
smaller than for full lattice QCD simulations. At the present stage of the project, simulations can
still be run on time scales of days on university PC clusters.
The drawback is that the effective action is
only valid in parameter ranges where the expansion converges, which is 
currently restricted to the heavy mass region and the confined phase. Even there,
the effective theory is unsuitable for long range correlation functions, but it
gives accurate results for bulk thermodynamic quantities and phase transitions \cite{test}.
In particular, it has already provided 
predictions with better than 10\% accuracy for the critical couplings of $SU(2),SU(3)$
Yang-Mills \cite{Langelage:2010yr}, the critical quark masses where the deconfinement transition 
changes to a crossover \cite{Fromm:2011qi} and the tricritical point of the deconfinement transition
at imaginary chemical potential \cite{pinke}. 
A similar approach is used in
\cite{Unger:2011it,Fromm:2011kq,Kawamoto:2005mq,Nakano:2010bg} 
with staggered fermions. 
There, the chiral regime can be studied directly but the strong coupling series is much harder to
compute and no continuum extrapolations are possible so far.

The paper is organised as follows. In section \ref{sec:efft} we summarise the derivation of the
effective action in the pure gauge sector and give a systematic calculation of the fermion determinant. In section \ref{sec:an} we analyse the effective action by analytic methods
to leading and next-to-leading order in the small effective couplings. Section \ref{sec:lang}
is devoted to a systematic study of the validity of complex Langevin simulations. Finally, section \ref{sec:phys} contains our physics results for the cold and dense 
regime of QCD with heavy quarks. Readers not interested in the technical aspects of the derivation
and simulation may skip sections \ref{sec:efft}, \ref{sec:lang} and read 
sections \ref{sec:an}, \ref{sec:phys} only.

\section{The effective action \label{sec:efft}}

Starting point is a $(3+1)$-dimensional lattice with Wilson's gauge and
fermion 
actions for $N_f$ flavours, which after Grassmann integration may be written as
\begin{eqnarray}
Z=\int[dU_\mu]\;\exp\left[-S_g\right]\prod_{f=1}^{N_f}\det\left[Q^f\right]\;,\qquad
-S_g=\frac{\beta}{2N_c}\sum_p\left[\tr\,   U_p+\tr\,   U_p^\dagger\right]\;,
\end{eqnarray}
with elementary plaquettes $U_p$, the quark hopping matrix for the flavour $f$,
\begin{eqnarray}
&&(Q^f)^{ab}_{\alpha\beta,xy}=\delta^{ab}\delta_{\alpha\beta}\delta_{xy}\\ \hspace*{-1.5cm}
&&-\kappa_f\sum_{\nu=0}^3\left[e^{a\mu_f\delta_{\nu0}}(1+\gamma_\nu)_{\alpha\beta}U_\nu^{ab}(x)
\delta_{x,y-\hat{\nu}}+e^{-a\mu_f\delta_{\nu0}}(1-\gamma_\nu)_{\alpha\beta}U_{-\nu}^{ab}(x)
\delta_{x,y+\hat{\nu}}\right]\;,
\;\nonumber
\end{eqnarray}
and $U_{-\nu}^{ab}(x) = U_{\nu}^{\dagger ab}(x-\hat{\nu})$. The effective action is then defined by integrating out the spatial link variables
\begin{eqnarray}
Z&=&\int[dU_0]\;\exp[-S_\mathrm{eff}]\;,\nn\\
\exp[-S_{\mathrm{eff}}]&\equiv&\int[dU_k]\exp\left[-S_g\right]\prod_{f=1}^{N_f}\det\left[Q^f\right]\;, \nn\\ S_{\mathrm{eff}} &=&  \sum_{i=0}^{\infty}S^s_i(\beta, \kappa_f,N_{\tau};W) + \sum_{i=1}^{\infty} S^a_i(\beta, N_{\tau}, \kappa_f, \mu_f;W)   \nn\\
&=& \sum_{i=0}^{\infty}S^g_{i}(\beta, \kappa_f,N_{\tau};W) + \sum_{i=0}^{\infty} S^f_{i}(\beta, N_{\tau}, \kappa_f, \mu_f;W) \;. 
\label{eq_defeffth}
\end{eqnarray}
In the first line we split into a part which is $Z(N_c)$ centre symmetric and a part with symmetry breaking terms. For the present work it is more convenient to split the action into a purely gluonic part 
and a fermionic part due to the determinant, which contains both symmetric and symmetry breaking
contributions.
All terms depend only 
on temporal Wilson lines $W_{\vec{x}}$ or their traces, the Polyakov loops,
\begin{eqnarray}
L_{\vec{x}}\equiv\tr\,   W_{\vec{x}}\equiv \tr\prod_{\tau=1}^{N_\tau}U_0\left(\vec{x},\tau\right)\;.
\end{eqnarray}
The effective action features an infinite tower of interaction
terms between loops to all powers and at all distances, where $S^{x}_{i}$ denote $i$-point-interactions.
These are completely determined in terms of Wilson lines and the parameters of the original theory.
Note that, without truncations, the effective action is unique and exact.
Non-perturbative determinations of the  effective action \cite{poly1,poly2,poly3,poly4,poly5} 
can in principle be applied 
at all parameter values. In practice they necessarily imply truncation and modelling,
which may have to be different in different parameter regimes. 
In our approach we compute the effective action in a combined strong coupling and
hopping parameter expansion, which orders terms according to their leading powers in $\beta, \kappa$.
By summing up all temporal windings we 
make sure that we 
have the complete dependence on chemical potential, or fugacity, in each order of the hopping
parameter  expansion.

\subsection{Pure gauge theory}

For the Yang-Mills part, it is advantageous to perform a character expansion
\begin{eqnarray}
\exp\left[\frac{\beta}{2N_c}\Big(\tr\,   U_p+\tr\,   U_p^\dagger\Big)\right]
=c_0(\beta)\left[1+\sum_{r\neq0}d_ra_r(\beta)\chi_r(U_p)\right]\;,
\end{eqnarray}
where the factor $c_0(\beta)$ can be neglected as it is independent of gauge links and
cancels in 
expectation values. In earlier publications
\cite{Fromm:2011qi,Langelage:2010yr,Langelage:2010nj}, we have shown how to compute
the effective 
gauge theory up to rather high orders in the fundamental character expansion
coefficient 
$u(\beta)\equiv a_f(\beta) = \frac{\beta}{18} + \ldots$ . In leading order we have a chain of $N_\tau$ fundamental 
plaquettes winding around the temporal direction and closing via periodic boundary
conditions. Therefore the leading order is a two-point interaction, 
\begin{eqnarray}
S^g_{2} (\beta, N_{\tau}) =\lambda(u,N_\tau)\sum_{<\vec{x} \vec{y}>}\left(L_{\vec{x}}L_{\vec{y}}^\ast+L_{\vec{x}}^\ast
L_{\vec{y}}\right)\;,
\qquad\lambda(u,N_\tau)=u^{N_\tau}\Big[1+\ldots\Big]\;,
\label{eq_seffgauge}
\end{eqnarray}
where higher order corrections of $\lambda(u,N_\tau)$ as well as a discussion of
higher order 
interaction terms can be found in \cite{Langelage:2010yr}. In the leading order
expression of eq.~(\ref{eq_seffgauge}) we already see that 
$\lambda(u,N_\tau)$ is suppressed for large $N_\tau$, since $u<1$, see also
\cite{Fromm:2011qi} 
for a further discussion. In this work we aim at temperatures $T\leq 10$ MeV with lattice 
parameters $\beta\lsim 6, N_\tau\geq 100$, where $\lambda\lsim 10^{-25}$ can be safely neglected.

\subsection{Static quark determinant}

The quark determinant is expanded in a hopping expansion. 
In order to keep the complete 
dependence on chemical potential it is useful to split the quark matrix in positive and negative temporal and spatial hops,
\begin{eqnarray}
Q=1-T-S=1-T^+-T^--S^+-S^-\;.
\end{eqnarray}
The static determinant is
then given 
by neglecting the spatial parts,
\begin{eqnarray}
\det[Q_{\mathrm{stat}}] &=& \det[1-T] = \det[1-T^+ - T^-] \nn \\
&=& \det \Big[1-\kappa e^{a\mu}(1+\gamma_0)U_0(x) \delta_{x,y-\hat{0}} \nn\\ &&\hspace*{1.2cm}
-\kappa e^{-a\mu}(1-\gamma_0)U^{\dagger}_0(x-\hat{0}) \delta_{x,y+\hat{0}}\Big]\;,
\end{eqnarray}
with propagation in the temporal direction only. Calculating the space and spin
determinant we get
\begin{eqnarray}
\det[Q_{\mathrm{stat}}]&=& \prod_{\vec{x}} 
\det \Big[1+(2 \kappa e^{a \mu})^{N_{\tau}}W_{\vec{x}}\Big]^2
\det \Big[1+(2 \kappa e^{-a \mu})^{N_{\tau}}W^{\dagger}_{\vec{x}}\Big]^2\;.
\label{q_static}
\end{eqnarray}
Note that this includes all windings of Wilson lines around the temporal direction and thus 
the full fugacity dependence.
A well-known
relation valid 
for $SU(3)$ then 
allows us to reformulate this in terms of Polyakov loops,
\begin{eqnarray}
\det[Q _{\mathrm{stat}}]&=& \prod_{\vec{x}} 
\left[1 + c L_{\vec{x}} + c^2 L^*_{\vec{x}}+c^3\right]^2
\left[1 + \bar{c} L^*_{\vec{x}} + \bar{c}^2 L_{\vec{x}}+\bar{c}^3\right]^2,
\label{eq_qsim}
\end{eqnarray}
with the abbreviation
\beq
c(\mu)\equiv\left(2\kappa e^{a\mu}\right)^{N_\tau}= e^{\frac{\mu-m}{T}} \equiv \bar{c}(-\mu)\;,
\eeq
and the constituent quark mass $am=-\ln(2\kappa)=\frac{am_B}{3}$, to leading  
order of eq. ($\ref{eq:hadron}$). When $\det[Q_{\mathrm{stat}}]$ is exponentiated, the parameter $c$ also constitutes the effective one-point coupling constant of $S^f_1$ to leading order \cite{Fromm:2011qi},
\beq
h_1=c, \quad \bar{h}_1=\bar{c}\;. 
\eeq

\subsection{Kinetic quark determinant}

In order to compute a systematic hopping expansion about the static limit, we define the kinetic quark
determinant
\begin{eqnarray}
\det[Q]&\equiv&\det[Q_{\mathrm{stat}}]\det[Q_{\mathrm{kin}}]\;,\nonumber\\
\det[Q_{\mathrm{kin}}]&=&\det[1-(1-T)^{-1}(S^++S^-)] \nonumber\\
&\equiv&\det[1-P-M]=\exp\left[\tr\,  \ln
(1-P-M)\right]\;,
\label{eq_detqkin}
\end{eqnarray}
which we then split into parts describing quarks moving in positive and negative
spatial 
directions, $P=\sum_kP_k$ and $M=\sum_kM_k$. The reason for this is that the trace
occurring 
in eq.~(\ref{eq_detqkin}) is also a trace in coordinate space. This means
that only closed loops contribute and hence
we need the same number of $P$s and $M$s in the expansion of the logarithm.
Through $\mathcal{O}\left(\kappa^4\right)$ we have
\begin{eqnarray}
\det[Q_{\mathrm{kin}}]&=&\exp\left[-\tr\,   PM-\tr\,   PPMM-
\frac{1}{2}\tr\,   PMPM\right]\left[1+\mathcal{O}(\kappa^6)\right] \\
&=&\left[1-\tr\,   PM - \tr\,   PPMM -
\frac{1}{2} \tr\,   PMPM+\frac{1}{2}\left(\tr\,   PM\right)^2\right]
\left[1+\mathcal{O}(\kappa^6)\right]\;. \nonumber
\label{eq_detqkin2}
\end{eqnarray}
The next step is to consider the different directions in $P$ and 
$M$ which also need to come in pairs, 
\begin{eqnarray}
\sum_{ij}\tr\,   P_iM_j&=&\sum_i\tr\,   P_iM_i\;,\label{eq_qdet2}\\
\sum_{ijkl}\tr\,  P_iP_jM_kM_l&=&\sum_{i}\tr\,  P_iP_iM_iM_i+
\sum_{i\neq j}\tr\,  P_iP_jM_iM_j \nonumber  \\
 && +\sum_{i\neq
j}\tr\,  P_iP_jM_jM_i\label{eq_ppmm}\;,\\
 \frac12 \sum_{ijkl}\tr\,  P_iM_jP_kM_l&=& \frac12 \sum_i \tr\,  P_iM_iP_iM_i+
 \frac12 \sum_{i\neq j}\tr\,  P_iM_iP_jM_j \nonumber  \\
 &&+ \frac12 \sum_{i\neq
j}\tr\,  P_iM_jP_jM_i\label{eq_pmpm}\;,\\
\frac12 \sum_{ijkl}\tr\,  P_iM_j\tr\,  P_iM_j&=&\frac12 \sum_{i,
j}\tr\,  P_iM_i\tr\,  P_jM_j \label{eq_trpm2}\;.
\end{eqnarray}

\subsection{Static quark propagator}

We now compute the static quark
propagator 
$(1-T)^{-1}$ appearing in eq.~(\ref{eq_detqkin}).
Since $(1+\gamma_\mu)(1-\gamma_\mu)=0$, hops in forward and backward time
direction 
do not mix and
 the full static quark propagator is given by
\begin{eqnarray}
(Q_{\mathrm{stat}})^{-1}=(Q^+_{\mathrm{stat}})^{-1}
+(Q^-_{\mathrm{stat}})^{-1}-1\;.
\end{eqnarray}
In order to compute the positive static quark propagator, we 
use the series expansion
\begin{eqnarray}
(Q^+_{\mathrm{stat}})^{-1}=\left(1-T^+\right)^{-1}=\sum_{n=0}^\infty
(T^+)^n\;.
\end{eqnarray}
The inverse is then given by
\begin{eqnarray}
(Q^+_{\mathrm{stat}})^{-1}_{\tau_1\tau_2} &=& \delta_{\tau_1\tau_2}\left(1-qz^{N_\tau}W\right) 
+qz^{\tau_2-\tau_1}W(\tau_1,\tau_2)\Big[\Theta(\tau_2-\tau_1)-z^{N_\tau}
\Theta(\tau_1-\tau_2)\Big]\;,\nn\\
q&\equiv&\frac{1}{2}(1+\gamma_0)\left(1+z^{N_\tau} W\right)^{-1}\;,\qquad z = 2\kappa e^{a\mu}\;.
\end{eqnarray}
$W(\tau_1,\tau_2)$ is a temporal Wilson line from $\tau_1$ to $\tau_2$ and we have suppressed its spatial
location. If 
$\tau_1=\tau_2$, the Wilson line winds around the lattice, $W(\tau_1,\tau_1)=W$. 
The contribution in negative time direction 
$(Q^-_{\mathrm{stat}})^{-1}_{\tau_1\tau_2}$ can then be obtained from 
$(Q^+_{\mathrm{stat}})^{-1}_{\tau_1\tau_2}$ by the following replacements
\begin{eqnarray}
\tau_1\leftrightarrow\tau_2\;,\qquad
W(\tau_1,\tau_2)\leftrightarrow W^\dagger(\tau_1,\tau_2)\;,\qquad
\mu\leftrightarrow-\mu\;,
\end{eqnarray}
and reads
\begin{eqnarray}
(Q^-_{\mathrm{stat}})^{-1}_{\tau_1\tau_2}&=&\delta_{\tau_1\tau_2}\left(1-\bar{q}\bar{z}^{N_\tau} W^\dagger
\right)+\bar{q}\bar{z}^{\tau_1-\tau_2}W^\dagger(\tau_1,\tau_2)\Big[\Theta(\tau_1-\tau_2)-\bar{z}^{N_\tau}
\Theta(\tau_2-\tau_1)\Big]\;, \nn \\
\bar{q}&=&\frac{1}{2}(1-\gamma_0)\left(1+\bar{z}^{N_\tau}W^\dagger\right)^{-1}\;,\qquad
\bar{z}=2\kappa e^{-a\mu}\;.
\end{eqnarray}
Finally we split the temporal quark propagator in spin space as well as in
propagation in positive 
and negative temporal direction according to
\beq
\label{eq_qstat}
\left(Q_{\mathrm{stat}}\right)^{-1}= A + \gamma_0 B = A^++\gamma_0B^+ + A^--\gamma_0 B^-\;,
\eeq
\begin{eqnarray}
A^+_{xy}&=&\frac12 \left[1-\frac{c W}{1+c W}\right]\delta_{xy}
+\frac{1}{2}z^{\tau_y-\tau_x}\frac{W(\tau_x,\tau_y)}{1+c W}\bigg[\Theta(\tau_y-\tau_x)-c
\Theta(\tau_x-\tau_y)\bigg]\delta_{\vec{x}\vec{y}}\;,\nonumber\\
B^+_{xy}&=&-\frac{1}{2}\frac{cW}{1+cW}\delta_{xy}
+\frac{1}{2}z^{\tau_y-\tau_x}\frac{W(\tau_x,\tau_y)}{1+cW}\bigg[\Theta(\tau_y-\tau_x)-c
\Theta(\tau_x-\tau_y)\bigg]\delta_{\vec{x}\vec{y}}\;,\nonumber\\
A^-_{xy}&=&\frac12 \left[1-\frac{\bar{c}W^\dagger}{1+\bar{c}W^\dagger}\right]\delta_{xy}
+\frac{1}{2}\bar{z}^{\tau_x-\tau_y}\frac{W^\dagger(\tau_x,\tau_y)}{1+\bar{c}W^\dagger}\bigg[\Theta(\tau_x-\tau_y)-\bar{c}
\Theta(\tau_y-\tau_x)\bigg]\delta_{\vec{x}\vec{y}}\;,\nonumber\\
B^-_{xy}&=&-\frac{1}{2}\frac{\bar{c}W^\dagger}{1+\bar{c}W^\dagger}\delta_{xy}
+\frac{1}{2}\bar{z}^{\tau_x-\tau_y}\frac{W^\dagger(\tau_x,\tau_y)}{1+\bar{c}W^\dagger}\bigg[\Theta(\tau_x-\tau_y)-\bar{c}
\Theta(\tau_y-\tau_x)\bigg]\delta_{\vec{x}\vec{y}}\;.\nonumber
\end{eqnarray}

\subsection{Gauge integrals for the leading fermionic action \label{sec:gi}}

Next we compute the leading strong coupling contribution to the effective action by performing the group integrations. We will arrange the fermionic part of the effective action as
\begin{eqnarray}
\int [dU_k] \prod_{f} \det[Q^f_{\mathrm{kin}}] = e^{\sum_{i = 1}^{\infty} S^f_{i}(\beta = 0, \kappa_f, N_{\tau},\mu_f)}\;.
\end{eqnarray}
Since it is not known how to analytically perform the gauge integral over links in the exponent, we have  expanded it in a Taylor series. After the integration we shall see that it is possible to resum some terms back into an exponential.
At the order $\kappa^4$ there are zero-point contributions (or vacuum graphs) from closed hops around a plaquette. 
In a strong coupling series these only contribute after being dressed with a plaquette, 
$\sim \kappa^4 u$, and thus are neglected here.  
The one-point contributions of the Polyakov 
loops constitute the static determinant and have been computed already.

\subsubsection{Two-point interaction}

Dealing with more than one trace, as in $\Big(\sum_{i}\tr\,  P_iM_i \Big)^2$,  it will be necessary to explicitly display spatial coordinates, i.e.
\begin{eqnarray}
(\tr\, P_i M_i)^2 =  \sum_{\vec{x},i} (\tr\, P_{\vec{x}, i}M_{\vec{x}, i}) \sum_{\vec{y},j} (\tr\, P_{\vec{y},j}M_{\vec{y},j})\;.
\label{PM^2}
\end{eqnarray}
Here we have to consider three different possibilities: The two nearest-neighbour 
contributions may share $0$, $1$ or $2$ sites, where only the last one contributes to the two-point interaction. 
To the order $\kappa^4$ it is then
\begin{eqnarray}
e^{-S^f_{2}} \equiv\int[dU_k] \Big[
-\sum_{i}\tr\, P_iM_i  
- \frac12 \sum_{i}\tr\,  P_iM_i P_i M_i \\
+ \frac12 \sum_{\vec{x},i} \tr\,  P_{\vec{x},i} M_{\vec{x},i} \tr\, P_{\vec{x},i} M_{\vec{x},i}
\Big]\;. \nn
\end{eqnarray}
The first contribution to the two-point interaction is of order $\kappa^2$:
\begin{eqnarray}
&&-\int[dU_k]\sum_{i}\tr\,  P_iM_i=
-\sum_i\int[dU_k]\tr\,  \Big[Q_{\mathrm{stat}}^{-1}\,S^+_i\,Q_{\mathrm{stat}}^{-1}\,
S^-_i\Big]\\
&=&-\frac{8 \kappa^2}{N_c}\sum_{u,i} \tr\,   B_{u,u}
\tr\,B_{u +\hat{\imath},u+\hat{\imath}} \nn \\
&=& -2h_2 \sum_{\vec{x},i} \Bigg[\bigg(\tr\,  
\frac{c W_{\vec{x}}}{1 + c W_{\vec{x}}} - \tr\, \frac{\bar{c}
W^{\dagger}_{\vec{x}}}{1 + \bar{c} 
W^{\dagger}_{\vec{x}}} \bigg)\bigg(
\tr\,   
\frac{c W_{\vec{x}+\hat{\imath}}}{1 + c W_{\vec{x}+\hat{\imath}}}
- \tr\,  \frac{\bar{c}  
W^{\dagger}_{\vec{x}+\hat{\imath}}}{1 + \bar{c}  W^{\dagger}_{\vec{x}+\hat{\imath}}}
\bigg) 
\Bigg] \;.\nonumber
\end{eqnarray}
Here we have used the expressions eq.~(\ref{eq_qstat}) for $B$,
evaluated the trace over 
spin and coordinate space and introduced the coupling 
\beq
h_2=\frac{\kappa^2N_\tau}{N_c}\;.
\eeq
The group integrations have been computed via
\begin{eqnarray}
\int dU \,U_{ij}U^\dagger_{kl}=\frac{1}{N_c}\delta_{il}\delta_{jk}\;.
\end{eqnarray}
Note that this enforces the spatial link variables to be at the same 
temporal location and yields a factor $N_\tau$ rather than $N_\tau^2$ 
from the two temporal traces. From now on we will skip the last step, where one 
has to insert the definitions of $A$ and $B$ and perform the temporal sums. Explicit expressions
for all types of terms appearing in the following can be found in the appendix.

The next correction to the two-point interaction is of order $\kappa^4$:
\begin{eqnarray}
&& -\frac12 \int[dU_k]\sum_{i}\tr\,  P_iM_i P_i M_i = \\
&&-\frac{16\kappa^4}{N_c^2}\sum_{u \neq v, i}\left[\tr\,
B_{u,v}B_{v,u}
\Big(\tr\, B_{u+\hat{\imath},u+\hat{\imath}}\Big)^2 +
\Big(\tr\,B_{u,u}\Big)^2 
\tr\,B_{u+\hat{\imath},v+\hat{\imath}}B_{v+\hat{\imath},u+\hat{\imath}} \right]\nn\\
&&-\frac{16\kappa^4}{(N_c^2 -
1)}\sum_{u, i}\bigg\lbrace\tr\,B_{u,u}B_{u,u}
\Big(\tr\, B_{u+\hat{\imath},u+\hat{\imath}}\Big)^2 +
\Big(\tr\,B_{u,u}\Big)^2 
\tr\,B_{u+\hat{\imath},u+\hat{\imath}}B_{u+\hat{\imath},u+\hat{\imath}}\nn\\
&&-\frac{1}{N_c}\left[\tr\,B_{u,u}B_{u,u} \tr\,B_{u +\hat{\imath}, u +\hat{\imath}}B_{u +\hat{\imath}, u +\hat{\imath}}
 + \Big(\tr\, B_{u,u}\Big)^2 \Big(\tr\,
B_{u+\hat{\imath},u+\hat{\imath}}\Big)^2\right]\bigg\rbrace\;.\nn \label{eq_det44}
\end{eqnarray}
In this calculation it can happen that there is a spatial
link which is occupied by four matrices and we need the group integral (see e.g.
\cite{Creutz:1978ub})
\\
\begin{eqnarray}
\int
dU\,U_{i_1j_1}U_{i_2j_2}U^\dagger_{k_1l_1}U^\dagger_{k_2l_2}&=&\frac{1}{N_c^2-1}\Big[\delta_{i_1l_1}\delta_{i_2l_2}\delta_{j_1k_1}\delta_{j_2k_2}+\delta_{i_1l_2}\delta_{i_2l_1}\delta_{j_1k_2}\delta_{j_2k_1}\Big]\\
&-&\frac{1}{N_c(N_c^2-1)}\Big[\delta_{i_1l_2}\delta_{i_2l_1}\delta_{j_1k_1}\delta_{j_2k_2}+\delta_{i_1l_1}\delta_{i_2l_2}\delta_{j_1k_2}\delta_{j_2k_1}\Big]\;.
\nonumber
\end{eqnarray}
\\
The next contribution of order $\kappa^4$ comes from eq.~(\ref{PM^2}), which is a two-point interaction in the case that $\vec{x} = \vec{y}$ and $i = j$: 
\begin{eqnarray}
&& \frac12 \int[dU_k]\sum_{\vec{x},i} \tr\,  P_{\vec{x},i} M_{\vec{x},i} \tr\, P_{\vec{x},i} M_{\vec{x},i}  \\
&=&\frac{32\kappa^4}{N_c^2}\sum_{u \neq v, i}\left[ \Big(\tr\, B_{u,u}\Big)^2 \Big(\tr\,
B_{v+\hat{\imath},v+\hat{\imath}}\Big)^2
+\tr\,B_{u,v}B_{v,u} \tr\,B_{u+\hat{\imath},v+\hat{\imath}}B_{v+\hat{\imath},u+\hat{\imath}}\right]\nn\\
&&+\frac{32\kappa^4}{N_c^2-1}\sum_{u,i}\Bigg\lbrace\Big(\tr\,
B_{u,u}\Big)^2 \Big(\tr\,
B_{u+\hat{\imath},u+\hat{\imath}}\Big)^2+\tr\,B_{u,u}B_{u,u}\tr\,B_{u+\hat{\imath},u+\hat{\imath}}
B_{u+\hat{\imath},u+\hat{\imath}}\nn\\
&&-\frac{1}{N_c}\bigg[\tr\,B_{u,u}B_{u,u}
\Big(\tr\, B_{u+\hat{\imath},u+\hat{\imath}}\Big)^2 +
\Big(\tr\,B_{u,u}\Big)^2 
\tr\,B_{u+\hat{\imath},u+\hat{\imath}}B_{u+\hat{\imath},u+\hat{\imath}}
\bigg]\Bigg\rbrace\;.\nn
\end{eqnarray}
Higher corrections to the two-point interaction start with $\mathcal{O}(\kappa^6)$.

\subsubsection{Three-point interaction}

The three-point interaction starts at $\mathcal{O}(\kappa^4)$;
\begin{eqnarray}
e^{-S^f_{3}} &\equiv&\int[dU_k] \Big[
-\sum_{i}\tr\, P_i P_i M_i M_i  
- \sum_{i \neq j}\tr\, P_i P_j M_j M_i \\
&&- \frac12 \sum_{i \neq j}\tr\, P_i M_i P_j M_j
- \frac12 \sum_{i \neq j}\tr\, P_i M_j P_j M_i 
+ \frac12 \sum_{\vec{x},\vec{y},i,j}  \tr\,  P_{\vec{x},i} M_{\vec{x},i} \tr\, P_{\vec{y},j} M_{\vec{y},j} 
\Big]\;. \nn
\end{eqnarray}
The different contributions are evaluated to be
\begin{align}
-\int[dU_k] \sum_{i}\tr\, P_i P_i M_i M_i  &=  \nonumber \\
-\frac{32 \kappa^4}{N_c^2} \sum_{u,v,i} &\tr\,
B_{u,u}\tr\,A_{u + \hat{\imath}, v + \hat{\imath}} A_{v + \hat{\imath}, u + \hat{\imath}} \tr\,B_{u + 2 \hat{\imath}, u + 2 \hat{\imath}}\;,
\end{align}
\begin{align}
-\int[dU_k] \sum_{i \neq j}\tr\, P_i P_j M_j M_i    &=  \nonumber \\
-\frac{16 \kappa^4}{N_c^2} \sum_{u,v,i \neq j} &\tr\,
B_{u-\hat{\imath},u-\hat{\imath}} \Big[\tr\,A_{u,v}A_{v,u}+\tr\,B_{u,v}B_{v,u}\Big]  \tr\,B_{u +  \hat{\jmath}, u + \hat{\jmath}}\;,
\end{align}
\begin{align}
 - \frac12 \int[dU_k] \sum_{i \neq j}\tr\, P_i M_i P_j M_j &=  \nonumber \\
 -\frac{8 \kappa^4}{N_c^2}\sum_{u,v,i \neq j}
&\tr\,B_{u+\hat{\imath},u+\hat{\imath}}
\Big[\tr\,A_{u,v}A_{v,u}+\tr\,B_{u,v}B_{v,u}\Big]
\tr\, B_{u+\hat{\jmath},u+\hat{\jmath}}\;,\\
-   \frac12 \int[dU_k] \sum_{i \neq j}\tr\, P_i M_j P_j M_i   &=  \nonumber \\
-\frac{8 \kappa^4}{N_c^2}\sum_{u,v,i \neq j}
&\tr\,B_{u-\hat{\imath},u-\hat{\imath}}
\Big[\tr\,A_{u,v}A_{v,u}+\tr\,B_{u,v}B_{v,u}\Big]
\tr\, B_{u-\hat{\jmath},u-\hat{\jmath}}\;,
\end{align}
\begin{align}
\frac12 \int[dU_k] \sum_{\vec{x}, \vec{y}, i, j}  \tr\,  P_{\vec{x},i} M_{\vec{x},i} \tr\, P_{\vec{y},j} M_{\vec{y},j}  &= \nonumber \\
 \frac{32\kappa^4}{N_c^2}\sum_{u,v,i,j}
 &\tr\,   B_{u,u} \tr\,B_{u+\hat{\imath},u+\hat{\imath}}\tr\,  
B_{v,v} \tr\,B_{v+\hat{\jmath},v+\hat{\jmath}}\;,
\end{align}
where the sum is only over terms where the two traces share one spatial point.

\subsubsection{Four-point interaction}

There are only two four-point interactions to order $\kappa^4$: 
\begin{eqnarray}
e^{-S^f_{4}} \equiv\int[dU_k] 
\Big[-\sum_{i \neq j}\tr\, P_i P_j M_i M_j + \frac12 \sum_{\vec{x},\vec{y},i,j} \tr\,  P_{\vec{x},i} M_{\vec{x},i} \tr\, P_{\vec{y},j} M_{\vec{y},j}\Big]\;.
\end{eqnarray}
After integration the first contribution vanishes in the strong coupling limit and only gives a non-zero contribution if a plaquette is inserted into the fermionic loop:
\begin{eqnarray}
\int[dU_k] \sum_{i \neq j}\tr\, P_i P_j M_i M_j  &=&\mathcal{O}(\kappa^4u)\;.
\end{eqnarray}
 Since we only calculate the action to order $\kappa^m u^n$ with $m+n=4$ we neglect this term.
The second contribution is
\begin{eqnarray}
 \frac12 \int[dU_k]\sum_{\vec{x},\vec{y},i,j} \tr\,  P_{\vec{x},i} M_{\vec{x},i} \tr\, P_{\vec{y},j} M_{\vec{y},j}  
 && = \\ 
 \frac{32\kappa^4}{N_c^2}\sum_{u,v, i , j}
 && \tr\,   B_{u,u} \tr\,B_{u+\hat{\imath},u+\hat{\imath}}\tr\,  
B_{v,v} \tr\,B_{v+\hat{\jmath},v+\hat{\jmath}}\;, \nn
\end{eqnarray}
where the sum is only over terms where the traces share no common spatial point.

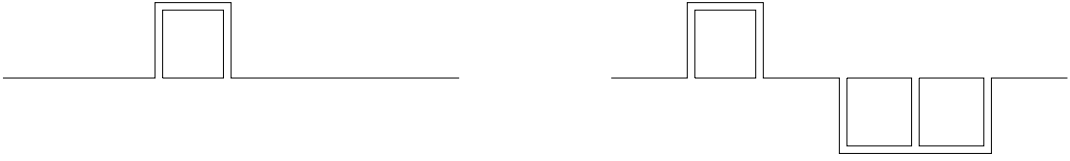
\begin{figure}
\scalebox{0.5}{
\begin{tikzpicture}
\draw(16,0) -- (20,0) -- (20,2) -- (22,2) -- (22,0) -- (28,0);
\draw(20.2,0) -- (20.2,1.8) -- (21.8,1.8) -- (21.8,0) -- (20.2,0);
\draw(32,0) -- (34,0) -- (34,2) -- (36,2) -- (36,0) -- (38,0) -- (38,-2) -- (42,-2)
-- (42,0) -- 
(44,0);
\draw(34.2,0) -- (34.2,1.8) -- (35.8,1.8) -- (35.8,0) -- (34.2,0);
\draw(38.2,0) -- (38.2,-1.8) -- (39.9,-1.8) -- (39.9,0) -- (38.2,0);
\draw(40.1,0) -- (40.1,-1.8) -- (41.8,-1.8) -- (41.8,0) -- (40.1,0);
\end{tikzpicture}
}
\caption{Finite gauge coupling corrections to the Polyakov line. After spatial link
integration 
these graphs give rise to terms $\sim \tr W$.}
\label{fig_pl}
\end{figure}

\subsection{Resummations}

In order to include as many terms as possible and improve convergence we perform a
resummation. Note that in order to perform the gauge integration, we had to expand the exponential
of hopping matrices, e.g.,
\begin{eqnarray}
e^{-\sum_i\tr P_i M_i} = 1 - \sum_i \tr P_i M_i + \frac12 \left(\sum_i \tr P_i M_i\right)^2 - \mathcal{O}(\kappa^6)\;.
\end{eqnarray}
After the integration it is possible to resum many of the resulting terms back into an exponential,
\begin{eqnarray}
\int[dU_k]  e^{-\sum_i\tr P_i M_i} = 1 
&-& \frac{8 \kappa^2}{N_c}\sum_{u,i} \tr\,   B_{u,u}
\tr\,B_{u +\hat{\imath},u+\hat{\imath}} \nn \\
&+& \frac{32\kappa^4}{N_c^2}\sum_{u,v, i,j}
 \tr\,   B_{u,u} \tr\,B_{u+\hat{\imath},u+\hat{\imath}}\tr\,  
B_{v,v} \tr\,B_{v+\hat{\jmath},v+\hat{\jmath}} \nn \\
= && e^{-\frac{8 \kappa^2}{N_c}\sum_{u,i} \tr\,   B_{u,u}
\tr\,B_{u +\hat{\imath},u+\hat{\imath}}} + \mathcal{O}(\kappa^6) \;.
\label{PM-resum}
\end{eqnarray}
Inspection of higher order terms indicates that this should always be possible.
Note that terms that have been resummed, like the higher orders in eq.~(\ref{PM-resum}), have to be excluded in the appropriate higher order to avoid double counting.

\subsection{Leading gauge corrections to the strong coupling limit}

Leaving the strong coupling limit, i.e.~for $\beta \neq 0$,  the gauge action 
has to be included when performing the group integration.
As a consequence the effective coupling constants depend on the gauge coupling also.
The leading gauge corrections are of order $N_{\tau} \kappa^2 u$ coming from
attaching plaquettes 
to the Wilson line, cf.~figure \ref{fig_pl}, and
\begin{eqnarray}
c\rightarrow 
h_1 =(2\kappa e^{a\mu})^{N_\tau} \ \Big[1+6\kappa^2 N_{\tau} u + \mathcal{O}(\kappa^2 u^5) \Big]\;.
\end{eqnarray}
This can also be exponentiated by summing over multiple attached, disconnected plaquettes at
different locations
\begin{eqnarray}
h_1&=& 
\exp\left[N_\tau\left(a\mu+\ln2\kappa+6 \kappa^2 \frac{u -
u^{N_{\tau}}}{1-u}\right)\right]\;,
\end{eqnarray}
and we see that in this way the Polyakov line receives mass corrections due to
interactions.
Note that this generates overcounting in higher orders, but in our opinion the 
resummation effects of this procedure more than compensates for this additional
care.
Let us finally  also give the gauge correction for the prefactor of the leading order of $S^f_{2}$
\begin{eqnarray}
h_2= \frac{\kappa^2N_\tau}{N_c}\left[1+2\frac{u-u^{N_\tau}}{1-u}+\ldots\right] \;.
\end{eqnarray}
This correction does not appear to exponentiate. 

\subsection{Effective action for the cold and dense regime}

The terms evaluated in the last sections and displayed in the appendix can now be added up to
provide the complete effective action. Fortunately, simplifications occur because some terms
have the same structure. Moreover, in this work we focus on the 
cold and dense regime and mostly simulate with $N_\tau > 100$, for which $\lambda\lsim 10^{-25}$, and terms that are of subleading order in $N_{\tau}$ as well as terms proportional to $\bar{h}_1$ 
are neglected, since $\bar{h}_1 \rightarrow 0$ as $T \rightarrow 0$. For $N_f=1$ we then simulate the simplified action
\begin{eqnarray}
\begin{split}
-S_{\mathrm{eff}}&= 
-\text{log} \sum_{\vec{x}} (1 + h_1 \text{Tr} W_{\vec{x}} + h_1^2 \text{Tr} W_{\vec{x}}^{\dagger} + h_1^3)^2 
-2 h_2 \sum_{\vec{x},i}  \text{Tr} \frac{h_1 W_{\vec{x}}}{1+h_1 W_{\vec{x}}} \text{Tr} \frac{h_1 W_{\vec{x}+i}}{1+h_1 W_{\vec{x}+i}} \\
&+ 2\frac{\kappa^4 N_{\tau}^2}{N_c^2} \sum_{\vec{x},i} \text{Tr} \frac{h_1 W_{\vec{x}}}{(1+h_1 W_{\vec{x}})^2} \text{Tr} \frac{h_1 W_{\vec{x}+i}}{(1+h_1 W_{\vec{x}+i})^2} \\
&+ \frac{\kappa^4 N_{\tau}^2}{N_c^2} \sum_{\vec{x},i,j}
\text{Tr} \frac{h_1 W_{\vec{x}}}{(1+h_1 W_{\vec{x}})^2} \text{Tr} \frac{h_1 W_{\vec{x}-i}}{1+h_1 W_{\vec{x}-i}}
\text{Tr} \frac{h_1 W_{\vec{x}-j}}{1+h_1 W_{\vec{x}-j}} \\
&+  2\frac{\kappa^4 N_{\tau}^2}{N_c^2} \sum_{\vec{x},i,j}
\text{Tr} \frac{h_1 W_{\vec{x}}}{(1+h_1 W_{\vec{x}})^2} \text{Tr} \frac{h_1 W_{\vec{x}-i}}{1+h_1 W_{\vec{x}-i}}
\text{Tr} \frac{h_1 W_{\vec{x}+j}}{1+h_1 W_{\vec{x}+j}} \\
&+ \frac{\kappa^4 N_{\tau}^2}{N_c^2} \sum_{\vec{x},i,j}
\text{Tr} \frac{h_1 W_{\vec{x}}}{(1+h_1 W_{\vec{x}})^2} \text{Tr} \frac{h_1 W_{\vec{x}+i}}{1+h_1 W_{\vec{x}+i}}
\text{Tr} \frac{h_1 W_{\vec{x}+j}}{1+h_1 W_{\vec{x}+j}} \;.
\end{split}
\end{eqnarray}
For $N_f=2$ some care has to be taken when introducing the determinant for the second flavour, which
introduces mixing terms that are not present in the above expression.

\subsection{Hadron masses in strong coupling and hopping expansion}

In order to interpret the results in the following sections,  it is convenient to also have the 
leading order of the meson and baryon masses,
\bea
am_M&=&-2\ln(2\kappa)-6\kappa^2-24\kappa^2\frac{u}{1-u}+\ldots\;,\nn\\
am_B&=&-3\ln (2\kappa)-18\kappa^2\frac{u}{1-u}+\ldots\;.
\label{eq:hadron}
\eea
To the orders given here, these expressions are the same for $N_f=1,2$ degenerate masses. 
From the second equation
we extract the running of the hopping parameter in the strong coupling limit for later use,
\beq
\left.\frac{\partial\kappa}{\partial a}\right|_{u=0}=-\kappa\frac{ m_B}{3}+O(\kappa^2)\;.
\label{eq:rk}
\eeq

\section{Analytic analysis of the effective theory \label{sec:an}}
 
\subsection{NLO perturbation theory for $N_f=1$}

A lot of  insight about the behaviour of the effective
theory can be gained by studying the static strong coupling limit, where the 
partition function factorises into a product of one-link integrals which can be solved
analytically. For the case of $N_f=1$ we previously observed the onset transition as a step function from zero density to lattice saturation \cite{Fromm:2012eb}. Here we extend this analysis beyond the
static strong coupling limit by using perturbation theory in the small couplings $\lambda,h_2$, 
permitting a clear understanding how the nuclear liquid gas transition is driven by 
interactions.  

To this end we consider the partition function with nearest-neighbour interaction between a Polyakov loop and its conjugate, as well as between two Polyakov loops, i.e.~including the couplings $\lambda,h_1,h_2$.
Here we are interested in the cold and dense regime. Near the zero temperature limit and for $\mu>0$, 
the anti-quark contributions vanish exponentially because  $\bar{h}_{1}\rightarrow 0 \quad \mbox{for} \quad T\rightarrow 0$ and the simplified partition function is
\bea
\label{zpt}
Z&=&\int[dW]\prod_{<\vec{x}, \vec{y}>}\left[1+\lambda(L_{\vec{x}}L_{\vec{y}}^*+L_{\vec{x}}^*L_{\vec{y}})\right]\prod_{\vec{x}}[1+h_1L_{\vec{x}}+h_2^2L_{\vec{x}}^*+h_1^3]^2
\\
&&\times \prod_{<\vec{x}, \vec{y}>}\left[1-2h_{2}\left(\frac{h_1L_{\vec{x}}+2h_1^2L_{\vec{x}}^*+3h_1^3}{1+h_1 L_{\vec{x}}+h_1^2L_{\vec{x}}^*+h_1^3}\right)\left(\frac{h_1L_{\vec{y}}+2h_1^2L_{\vec{y}}^*+3h_1^3}{1+h_1L_{\vec{y}}+h_1^2L_{\vec{y}}^*+h_1^3}\right)\right]\;.\nn
\eea
Note that the coupling $h_1$ parametrises $(\mu-m)$ and moreover approaches one around the onset transition. Therefore it cannot serve as an 
expansion parameter. On the other hand, there are physically interesting parameter regimes where 
$\lambda,h_2$ are sufficiently small to allow
for a power series expansion. 
The leading orders for the partition function and pressure read
\bea
Z&=&z_0^{N_s^3}+6\lambda N_s^3z_0^{N_s^3-2}z_1z_2-6h_2N_s^3z_0^{N_s^3-2}z_3^2\;,\nn\\
p&=&\frac{T}{V}\ln Z=\frac{1}{N_\tau N_s^3}\ln Z\nn\\
&=&N_\tau^{-1}\left(\ln z_0+6\lambda\frac{z_1z_2}{z_0^2}-6h_2\frac{z_3^2}{z_0^2}\right)\;,
\eea
with
\begin{eqnarray}
z_0&=&1+4h_1^3+h_1^6\;,\nn\\
z_1&=&3h_1^2+2h_1^5\;,\nn\\
z_2&=&2h_1+3h_1^4\;,\nn\\
z_3&=&6h_1^3+3h_1^6\;.
\end{eqnarray}
In the cold and dense regime we are working with $N_\tau\geq 10$ for which 
$\lambda(\beta=6.0,N_\tau)<10^{-5}$ plays no quantitative role, so we neglect it from here on.
The static strong coupling limit is obtained for $\lambda=h_2=0$ and has already been discussed in 
\cite{Fromm:2012eb}. In this case the partition function factorises into one-link partition functions $z_0$, i.e.~it represents a non-interacting system. We identify $z_0$ to consist of baryons, a spin 
3/2 quadruplet and a spin 0 baryon made of six quarks. Note that the Pauli principle
for $N_f=1$ does not admit spin 1/2 doublets. 
The quark number density and the energy density then follow as
\begin{eqnarray}
a^3n&=&\frac{1}{N_\tau N_s^3}\frac{\partial}{\partial a\mu}\ln Z\nn\\
&=&\frac{1}{N_\tau N_s^3}\frac{\partial h_1}{\partial a\mu}\frac{\partial}{\partial h_1}\ln Z\nonumber\\
&=&\frac{12h_1^3+6h_1^6}{z_0}-648h_2\frac{h_1^6(2+h_1^3)(1+h_1^3+h_1^6)}{z_0^3}\nonumber\\
&=&3a^3n_B\;,
\label{eq:density}
\end{eqnarray}
\begin{eqnarray}
a^4e&=&-\frac{a}{N_\tau N_s^3}\frac{\partial}{\partial a}\ln Z\bigg\vert_z\nonumber\\
&=&-\frac{a}{N_\tau N_s^3}\left(\frac{\partial h_1}{\partial a}\right)\bigg\vert_z\frac{\partial}{\partial h_1}\ln Z+\frac{6a}{N_\tau}\left(\frac{\partial h_2}{\partial a}\right)\left(\frac{z_3}{z_0}\right)^2\nn\\
&=&am_Ba^3n_B-4am_B\frac{h_2}{N_\tau}\left(\frac{z_3}{z_0}\right)^2\;,
\label{eq:e-density}
\end{eqnarray}
where we have made use of eq.~(\ref{eq:rk}).

\subsection{The nuclear liquid gas transition for $N_f=1$ \label{sec:pt}}

\begin{figure}[t]
\centerline{
\includegraphics[width=0.5\textwidth]{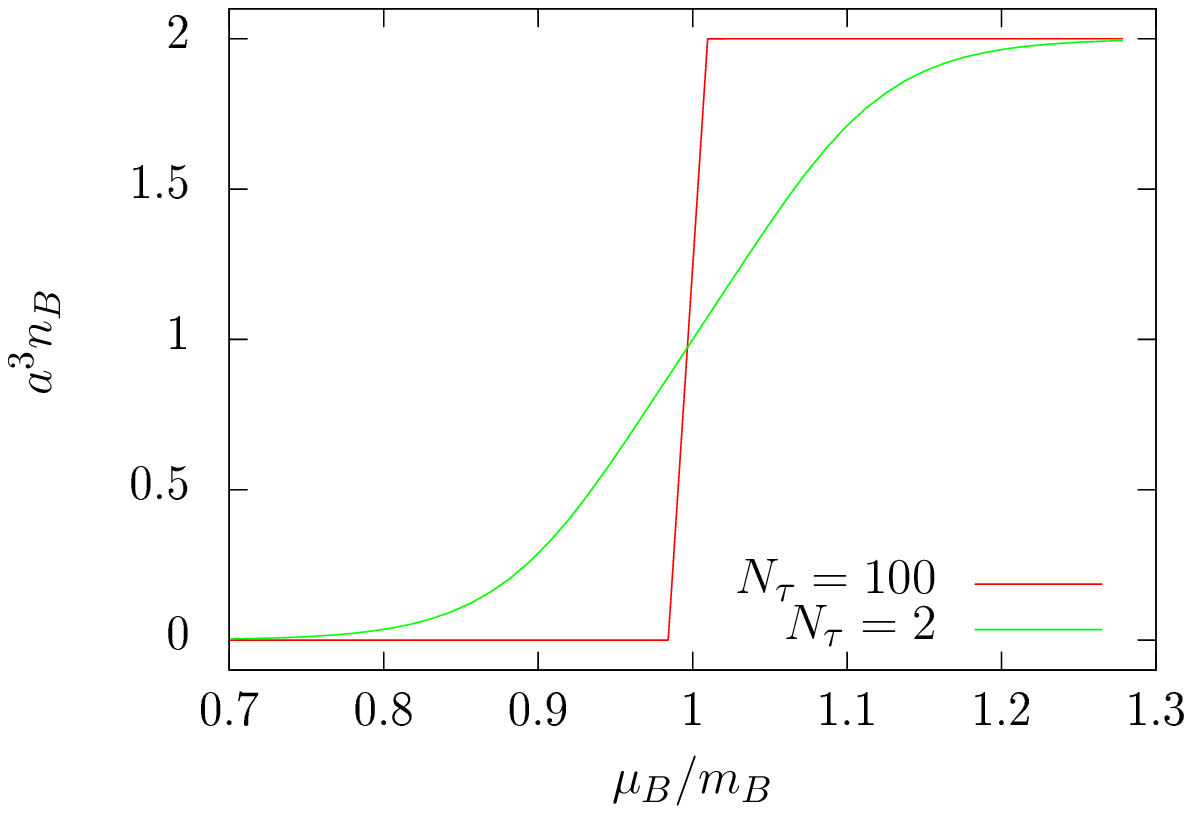}
\includegraphics[width=0.5\textwidth]{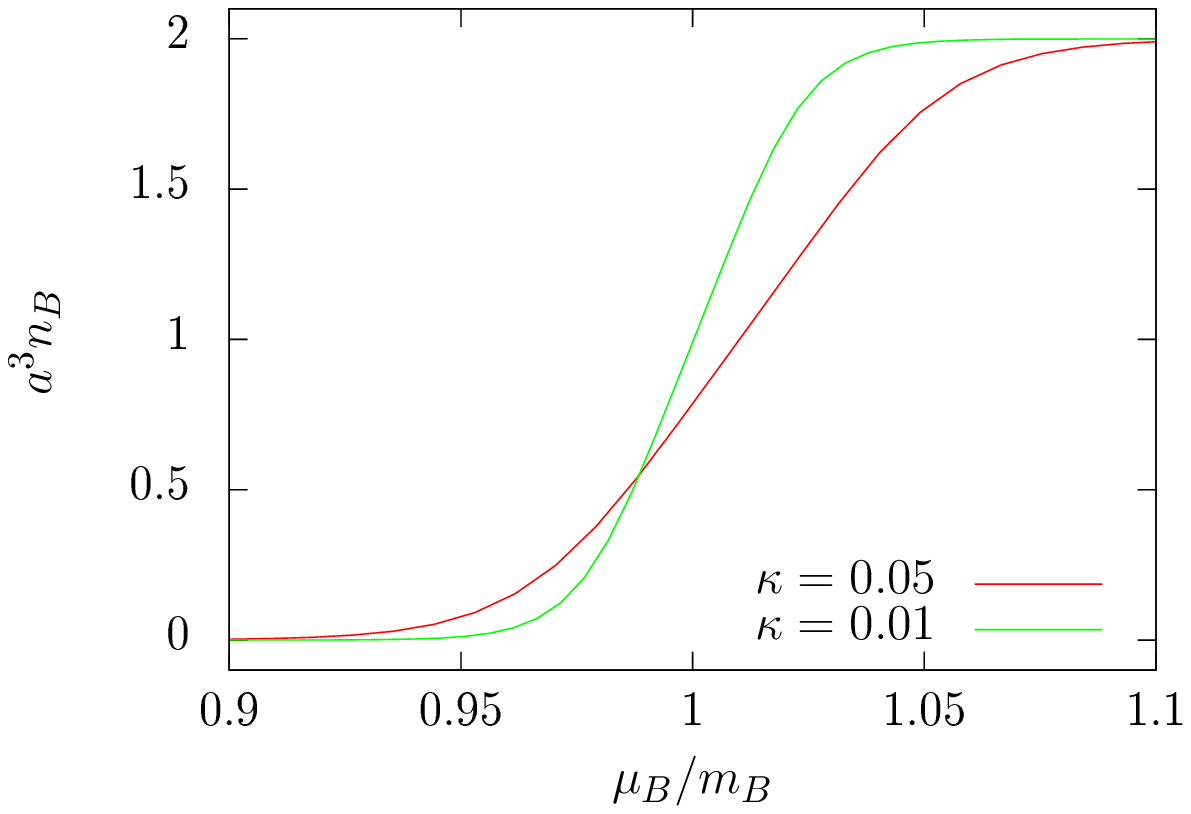}
}
\caption[]{The onset transition in lattice units, eq.~(\ref{eq:density}), for $\kappa=0.01,\beta=0$ and different $N_\tau$ (left) and for $N_\tau=10,\beta=0$ and different $\kappa$ (right).}
\label{fig:onset}
\end{figure}

With these formulae at hand, it is easy to analyse the physics of the cold and dense regime. Let us
begin with the static strong coupling limit.
At high density, the lattice is populated until it is saturated with six quarks per lattice
site according to the Pauli principle,
\beq
\lim_{\mu\rightarrow\infty}(a^3n)=2\cdot  N_c \equiv 2(a^3n_{B,\mathrm{sat}})\;.
\eeq 
Note that the dominating contribution to $z_0$ is a bosonic
baryon. However, it is a composite of quarks such that the Pauli principle,
built into the partition function in the original QCD action, is still contained in $z_0$.  
Another limit of interest is that of zero temperature. In this case we have
\bea
\lim_{T\rightarrow 0} a^4p&=&\left\{\begin{array}{cc} 0, & \mu<m\\
	2N_c (a\mu-am), & \mu>m\end{array}\right.\;,\nn\\
\nn\\
\lim_{T\rightarrow 0} a^3n&=&\left\{\begin{array}{cc} 0, & \mu<m\\
	2N_c, & \mu>m\end{array}\right.\;.
\eea
Thus we find the so-called silver blaze property, i.e.~the thermodynamic functions stay zero as the
chemical potential is raised until it crosses the constituent quark mass. Then it is possible to excite 
baryons and the onset phase transition to nuclear matter takes place. In the static strong coupling limit, 
this transition is a step function from zero to saturation density. This step function gets immediately 
smeared out to a smooth 
crossover as soon as a finite temperature ($N_\tau<\infty$) or coupling $h_2$ is switched on, cf.~figure \ref{fig:onset}.

We can unambiguously identify this transition as baryon condensation by also looking at the energy
density. Away from the static limit, there are non-vanishing attractive quark-quark (and hence 
baryon-baryon) interactions parametrised by $h_2$. These are identified by the quantity
\beq
\epsilon\equiv\frac{e-n_Bm_B}{n_Bm_B}=\frac{e}{n_Bm_B}-1\;,
\label{eq:bind}
\eeq
which gives the energy per baryon minus its rest mass in units of $m_B$. 
For temperatures approaching zero,
this is the binding energy per baryon. 
In perturbation theory, the result is
\beq
\epsilon=-\frac{4}{3}\frac{1}{a^3n_B}\left(\frac{z_3}{z_0}\right)^2\,\kappa^2=
-\frac{1}{3}\frac{1}{a^3n_B}\left(\frac{z_3}{z_0}\right)^2\, e^{-am_M}\;,
\label{eq:bindpt}
\eeq
where we have used the leading order of eq.~(\ref{eq:hadron}) to express the hopping parameter
by the meson mass. This result beautifully illustrates several interesting physics points.
In the non-interacting static limit with $\kappa=h_2=0$, there is no binding energy and hence no
true phase transition for the onset to nuclear matter. For finite $\kappa$ we see from the behaviour 
of $z_3,z_0$ that for $\mu<m$ and $T\rightarrow 0$ the binding energy is also zero, 
another manifestation of the silver blaze phenomenon. On the other hand, for $\mu>m, T\rightarrow 0$ 
it is explicitly negative and its absolute value increases with decreasing meson mass. 
This is in complete accord with expectations from nuclear physics models based on meson exchange. 

\begin{figure}[t]
\centerline{
\includegraphics[width=0.5\textwidth]{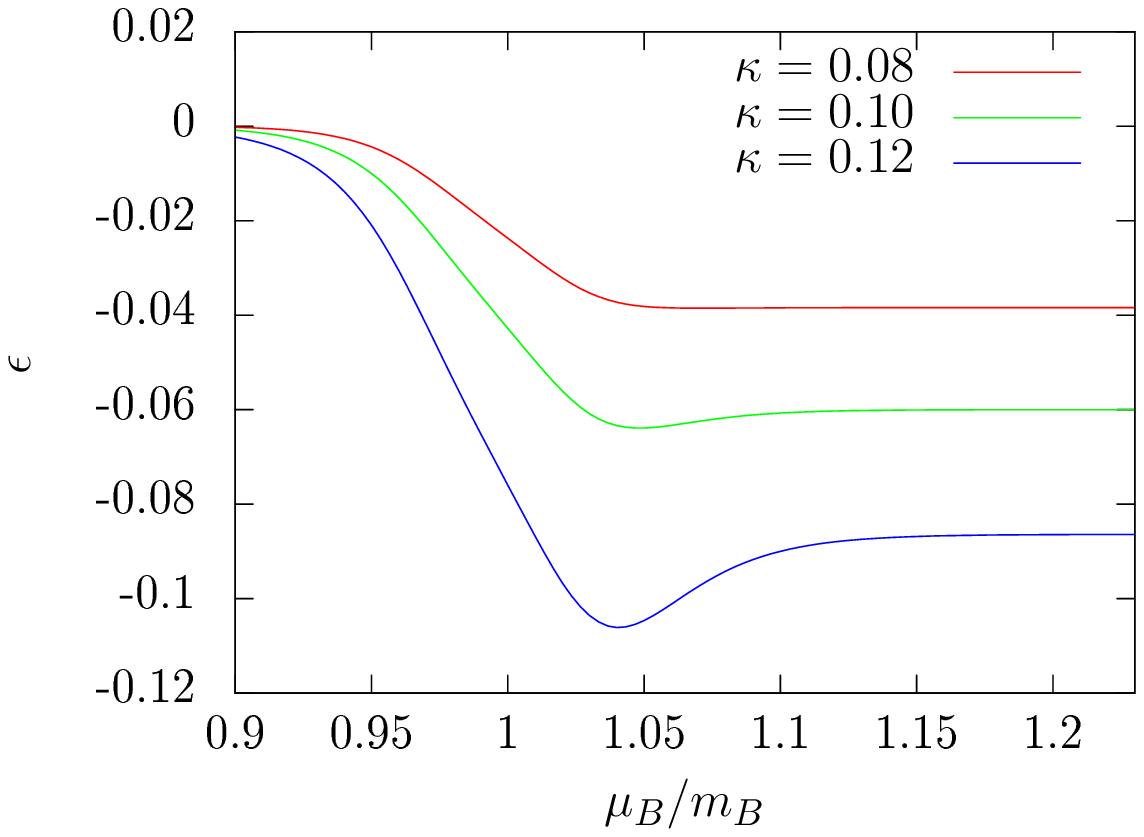}
\includegraphics[width=0.5\textwidth]{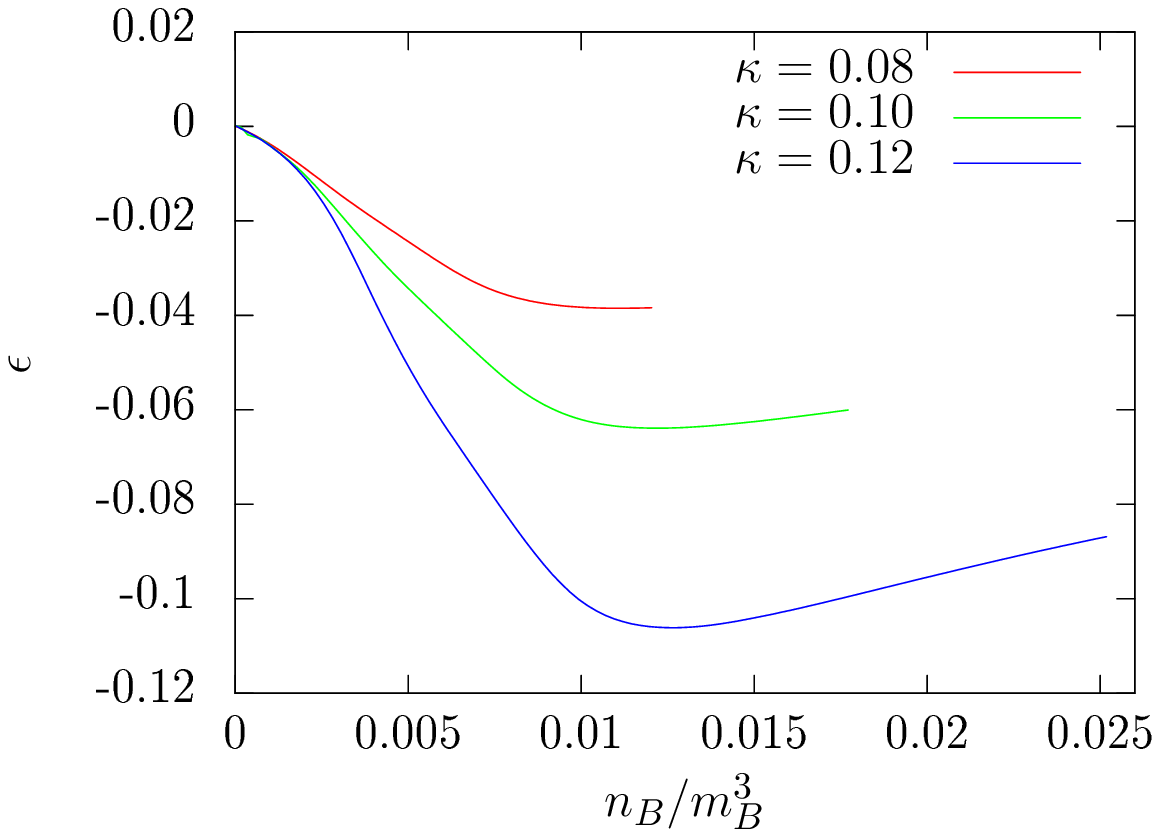}
}
\caption[]{Binding energy per nucleon in the strong coupling limit, eq.~(\ref{eq:bindpt}) with $N_\tau=10$. Quark mass decreases with growing $\kappa$.}
\label{fig:binda}
\end{figure}

The binding energy as a function of chemical potential is shown in figure \ref{fig:binda} (left), the 
asymptotes towards larger chemical potential are due to lattice saturation.
Plotting against the number density, we obtain the equation of
state as qualitatively expected for nuclear matter, figure \ref{fig:binda} (right). 
In particular, the binding energy per baryon gets more negative
as the quarks get lighter, until we see a minimum forming. Note that all curves eventually should turn upwards again, but for finite lattice spacing they are limited by the saturation density. With the explicit 
mass dependence in the binding energy and without a continuum extrapolation,
quantitative predictions for physical QCD cannot be made until the physical
flavour content and masses can be controlled. Nevertheless, it is interesting to keep in mind 
the physical binding energy per nucleon, $\epsilon\approx 0.017$ and the nuclear saturation density, 
$n_{B0}/m_\text{proton}^3\approx 0.016$.

\subsection{The static strong coupling limit for $N_f=2$ at finite baryon density}

For $\beta=0$, the partition function consists of the static determinant factors only
\begin{eqnarray}
Z (\beta=0)= \Big[ \int &[dW]& \prod_{\vec{x}} (1 + h_u L_{\vec{x}} + h_u^2 L_{\vec{x}}^{*} + h_u^3)^2 (1 + \bar{h}_u L_{\vec{x}}^{*} + \bar{h}_u^2 L_{\vec{x}} + \bar{h}_u^3)^2   \\
&&(1 + h_d L_{\vec{x}} + h_d^2 L_{\vec{x}}^{*} + h_d^3)^2 (1 + \bar{h}_d L_{\vec{x}}^{*} + \bar{h}_d^2 L_{\vec{x}} + \bar{h}_d^3)^2\Big]^V = z_0^V \;.\nn
\label{eq:ssc}
\end{eqnarray}
We again consider the zero temperature limit at $\mu>0$, for which
the anti-quark contributions vanish.
After the gauge integration the result reads
\begin{eqnarray}
z_0& =& (1 + 4 h_d^3 + h_d^6)+ (6 h_d^2 + 4 h_d^5) h_u+ (6 h_d + 10 h_d^4)h_u^2+ 
  (4 + 20 h_d^3 + 4 h_d^6)h_u^3 \nn \\
&&  + (10 h_d^2 + 6 h_d^5) h_u^4+ ( 4 h_d + 6 h_d^4) h_u^5 
  +(1 + 4 h_d^3 + h_d^6)h_u^6\;.
\label{eq:freegas}  
\end{eqnarray}
All exponents of $h_u^nh_d^m$ come in multiples of three, $n+m=\rm{mod} \;3$.
Just as in the one-flavour case (with $h_d=0$), this has the form of a free baryon gas where the prefactors
give the degeneracy of the spin multiplets.  Note that for $N_f=2$ we also find the standard spin 1/2
nucleons and many more combinations. 
To illustrate the prefactors, consider the example $h_u^2h_d$. There is the 
spin 1/2 doublet, the proton, as well as a spin 3/2 quadruplet, the $\Delta^+$, i.e.~six states altogether.
The states corresponding to $h_d^2h_u$ are the neutron and the $\Delta^0$, while  
$h_u^3,h_d^3$  are the $\Delta^{++},\Delta^-$ quadruplets, respectively. 
It continues with six-quark states. For example, $h_u^4h_d^2$ has
10 allowed spin-flavour combinations, corresponding to three spin 1 triplets and one spin 0 singlet.
These are consistent with an interpretation as di-baryon states built of $\Delta^{++}\Delta^0$ or $pp$.
Thus, eq.~(\ref{eq:freegas}) contains all baryonic spin-flavour multiplets that are consistent with the Pauli principle, i.e.~up to a
maximum of 12 constituent quarks.

The quark density reads
\bea
n_B&=&\frac{T}{V}\frac{\partial}{\partial \mu_B}\ln Z \nn\\
&=&2\Big[h_u^3 (2 + h_u^3) + h_d h_u^2 (3 + 4 h_u^3) + h_d^5 h_u (4 + 9 h_u^3) \nn \\
&&+  h_d^4 h_u^2 (10 + 9 h_u^3) + h_d^2 h_u (3 + 10 h_u^3)  \nn\\
&& +  h_d^6  (1 + 6 h_u^3 + 2 h_u^6) + h_d^3 (2 + 20 h_u^3 + 6 h_u^6)\Big]\nn \\
 &/&
     \Big[1 + 
   4 h_u^3 + h_u^6 + 2 h_d^4 h_u^2 (5 + 3 h_u^3) + 2 h_d^2 h_u (3 + 5 h_u^3) 
+ h_d^5 (4 h_u + 6 h_u^4) \nn\\ 
&&+ h_d (6 h_u^2 + 4 h_u^5) + 
   h_d^6 (1 + 4 h_u^3 + h_u^6) + 4 h_d^3 (1 + 5 h_u^3 + h_u^6)\Big]  \;.
\eea
In the high density limit numerator and denominator are dominated by the term with the highest power
and we obtain
\beq
\lim_{\mu\rightarrow\infty}(a^3n)=2\cdot 2\cdot N_c \equiv 4(a^3n_{B,\mathrm{sat}})\;.
\eeq 
This is the saturation density with two spin, two flavour
and three colour states per lattice site and fermion. 
In the zero temperature limit we have again the silver blaze property followed by 
a transition to lattice saturation
\bea
\lim_{T\rightarrow 0} a^4p&=&\left\{\begin{array}{cc} 0, & \mu<m\\
	4N_c (a\mu-am), & \mu>m\end{array}\right.\;,\nn\\
\nn\\
\lim_{T\rightarrow 0} a^3n&=&\left\{\begin{array}{cc} 0, & \mu<m\\
	4N_c, & \mu>m\end{array}\right.\;.
\eea
  
\subsection{The static strong coupling limit for $N_f=2$ at finite isospin density \label{sec:iso}}

Finite isospin density is realised for $\mu_I=\mu_u=-\mu_d$ \cite{son}. Choosing $\mu_u>0$, the zero temperature limit implies
$
\bar{h}_u,h_d\rightarrow 0 \quad \mbox{for} \quad T\rightarrow 0.
$
Omitting the corresponding terms from eq.~(\ref{eq:ssc}) and performing the gauge integration we
find the expression
\begin{eqnarray}
z_0 &=& (1 + 4 \bar{h}_d^3 + 
   \bar{h}_d^6) + (4 \bar{h}_d + 6 \bar{h}_d^4) h_u + (10 \bar{h}_d^2 + 6 \bar{h}_d^5) h_u^2 + (4 + 
    20 \bar{h}_d^3 + 4 \bar{h}_d^6) h_u^3 \nn\\ 
 &&   + (6 \bar{h}_d + 10 \bar{h}_d^4) h_u^4 + (6 \bar{h}_d^2 + 
    4 \bar{h}_d^5) h_u^5 + (1 + 4 \bar{h}_d^3 + \bar{h}_d^6) h_u^6\;.
\end{eqnarray} 
With isospin chemical potential, $d$-anti-quarks are now playing the same role as $u$-quarks
and the partition function is a free gas of baryons, anti-baryons and mesons. 
Differentiating with respect to isospin chemical potential gives the isospin density,
\bea
n_I&=&\frac{T}{V}\frac{\partial}{\partial \mu_I}\ln Z\\
&=&2 \Big[3 h_u^3 (2 + h_u^3) + 5 \bar{h}_d^4 h_u (3 + 8 h_u^3) + \bar{h}_d h_u (4 + 15 h_u^3) 
+\bar{h}_d^5 h_u^2 (21 + 20 h_u^3) \nn\\
&&+ \bar{h}_d^2 h_u^2 (20 + 21 h_u^3) + 3 \bar{h}_d^6 (1 + 6 h_u^3 + 2 h_u^6)  
   + 6 \bar{h}_d^3 (1 + 10 h_u^3 + 3 h_u^6)\Big]\nn\\
   &/&\Big[1 + 4 h_u^3 + h_u^6 + 
   2 \bar{h}_d^2 h_u^2 (5 + 3 h_u^3) + 2 \bar{h}_d^4 h_u (3 + 5 h_u^3) \nn\\
 &&  + 
   \bar{h}_d (4 h_u + 6 h_u^4) + \bar{h}_d^5 (6 h_u^2 + 4 h_u^5) + 
   \bar{h}_d^6 (1 + 4 h_u^3 + h_u^6) + 4 \bar{h}_d^3 (1 + 5 h_u^3 + h_u^6)\Big]\;. \nn
\eea
Also in this case, we find saturation in the high density limit,
\beq
\lim_{\mu\rightarrow\infty}(a^3n_I)=2\cdot 2\cdot N_c \equiv 4(a^3n_{I,\mathrm{sat}})\;.
\eeq 
Just as in the case of finite baryon density, the high density expression is dominated by a bosonic 
composite state which "knows" that it consists of constituent quarks, of which only a finite number can
be packed on one lattice site. The saturation level is hence identical to that for 
large baryon chemical potential.

Similarly, in the zero temperature limit we find again the silver blaze property followed by a non-analytic
transition to isospin condensation,
\bea
\lim_{T\rightarrow 0} a^4p&=&\left\{\begin{array}{cc} 0, & \mu<m\\
	4N_c (a\mu-am), & \mu>m\end{array}\right.\;,\nn\\
\nn\\
\lim_{T\rightarrow 0} a^3n_I&=&\left\{\begin{array}{cc} 0, & \mu<m\\
	4N_c, & \mu>m\end{array}\right.\;.
\eea
Note that for static quarks, $m_B/3= m_\pi/2$ and the onset transition to nuclear or isospin matter fall on top
of each other as a function of quark chemical potential. We shall see in our numerical investigations that
a gap between them opens up as expected when interactions between the hadrons are switched on.

\section{Simulation of the effective theory by complex Langevin \label{sec:lang}}

The effective theory specified in the last sections has a sign problem. With less
degrees of freedom 
and the theory being only three-dimensional, the sign problem is milder than in the
original theory
such that Monte Carlo methods are feasible at finite temperatures and chemical
potentials $\mu/T\lsim 3$ \cite{Fromm:2011qi}.
If, however, one is interested in cold dense matter in the zero
temperature limit, the sign problem becomes strong and Monte Carlo methods are
restricted to small volumes.
Fortunately, the effective theory is amenable to simulations using complex Langevin 
algorithms (for an introductory review, see \cite{dh}) and the onset transition to
nuclear matter could be demonstrated explicitly for 
very heavy quarks \cite{Fromm:2012eb}. In this section we discuss the validity of
complex Langevin for the effective 
theory.  We will only sketch the general method here, as there is an abundant
literature on this subject 
\cite{dh,clsu3,bilic88,etiology,su3lang}.

The basic idea is to introduce a fictitious Langevin time $\theta$, in which a field
theoretical 
system with a generic field $\phi$  evolves according to the Langevin equation
\beq
\label{langevin-eq}
\frac{\partial \phi(x,\theta)}{\partial \theta}=-\frac{\delta S}{\delta
\phi(x,\theta)}+\eta(x,\theta)\;,
\eeq
where $\eta(x,\theta)$ denotes Gaussian noise. 
In the case of a complex action, the field variables have to be complexified too, 
$\phi\rightarrow \phi_r + i\phi_i$. 
In our case, the degrees of freedom of the effective theory are
the  temporal Wilson lines
\beq
\int [d U_0]  f(W,W^\dag) = \int [d W]
f(W,W^\dag)\;.
\eeq
We may further simplify this by taking the trace of the Wilson lines and
parametrising the resulting Polyakov loops in terms of two
angles,
bringing them into a diagonal form \cite{gross83}
\beq
L(\theta,\phi) = e^{i \theta}+e^{i \phi}+e^{-i (\theta+\phi)}\;.
\eeq
This introduces a potential term denoted by $e^V$ with
\beq
V=\frac12  \mathrm{ln}(27-18|L|^2+8 \mathrm{Re}(L^3)-|L|^4)\;.
\eeq
Hence the integration measure we use in our simulation is the reduced Haar measure
\beq
\int [d W] = \int [dL] e^V = \int_{-\pi}^\pi [d\theta] \int_{-\pi}^\pi [d\phi] \ e^{2V}\;.
\eeq
This means instead of an integration over SU(3) matrices we have 2
complex degrees of freedom on every spatial lattice point. 
Furthermore, having only diagonal matrices their inversion 
is trivial.
With these ingredients eq.(\ref{langevin-eq}) was solved numerically using stepsizes
of around $\epsilon = 10^{-3}$ and applying the adaptive stepsize technique proposed
in \cite{adaptive-stepsize} to avoid numerical instabilities. \\

\begin{figure}[t]
\centerline{
\includegraphics[width=0.5\textwidth]{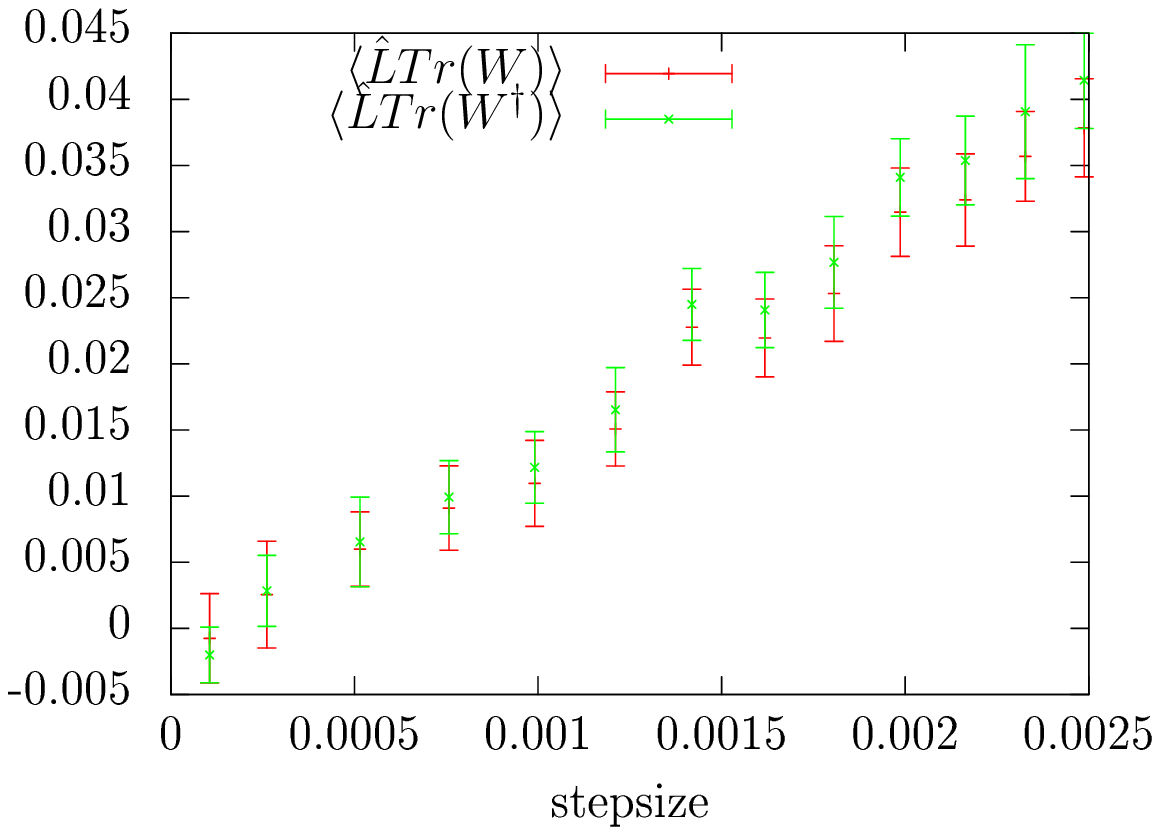}
\includegraphics[width=0.5\textwidth]{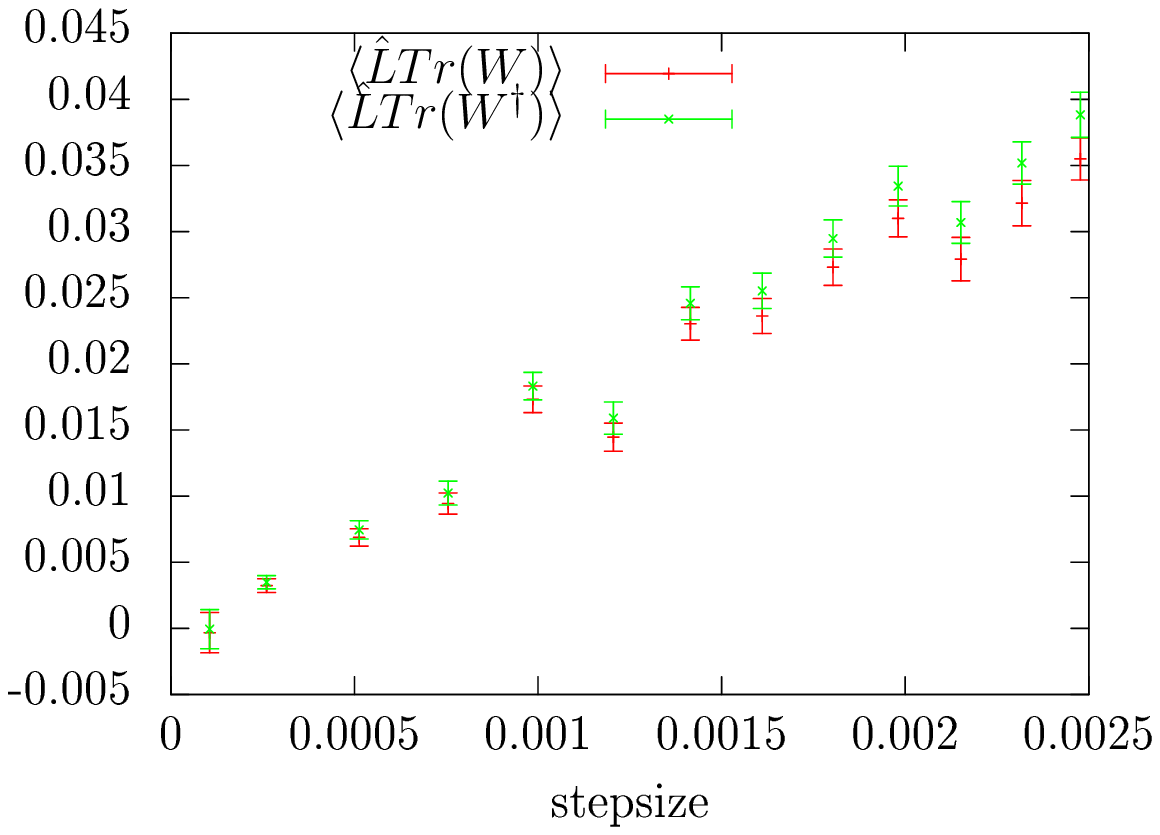}
}
\caption[]{Test of the convergence criterion for complex Langevin in the effective
theory to order
$\kappa^2$ (left) and $\kappa^4$ (right) for $\kappa^2 N_{\tau}/N_c = 0.01 $ and $\beta = 5.7$. 
$\hat{L}$ refers to the Langevin operator in (\ref{eq:lop})}
\label{fig:convcrit}
\end{figure}

\subsection{Criteria for correctness}

It is well known that the complex Langevin algorithm is not a general
solution to the 
complex action problem since it converges to the wrong limit in some cases, including
some parameter ranges for QCD \cite{dh,amb86}. The failure can be attributed to
insufficient localisation of
the probability distribution in the complex field space, and a set of criteria was
developed 
to check whether this localisation is sufficient in a given simulation \cite{etiology}. 
A necessary condition is that
the expectation value of all observables $O[\phi]$ vanishes after a Langevin operator
$\hat{L}$ has been 
applied to them,
\beq
\langle \hat{L}O[\phi]\rangle=0, \quad
\hat{L}=\sum_{a,x}\left(\frac{\partial}{\partial \phi_a(x)}
-\frac{\partial S}{\partial \phi_a(x)}\right)\frac{\partial}{\partial \phi_a(x)}\;.
\label{eq:lop}
 \eeq
 While, strictly speaking, this test is necessary on {\it all} observables of the
theory, in practice only
 a select few can be tested. Note that in the framework of our effective theory, all observables
 are expressed as functions of Polyakov loops and one might hope that its proper behaviour 
 propagates to more complicated functions of it. In figure \ref{fig:convcrit} we show the expectation
value of the Polyakov loop as a function of the step size of the Langevin algorithm 
for the effective theory to order $\kappa^2$ (left) and $\kappa^4$ (right). 
In both cases the criterion is satisfied in the limit of vanishing stepsize.  

\subsection{The logarithm of the static determinant}

\begin{figure}[t]
\centerline{
\includegraphics[width=0.5\textwidth]{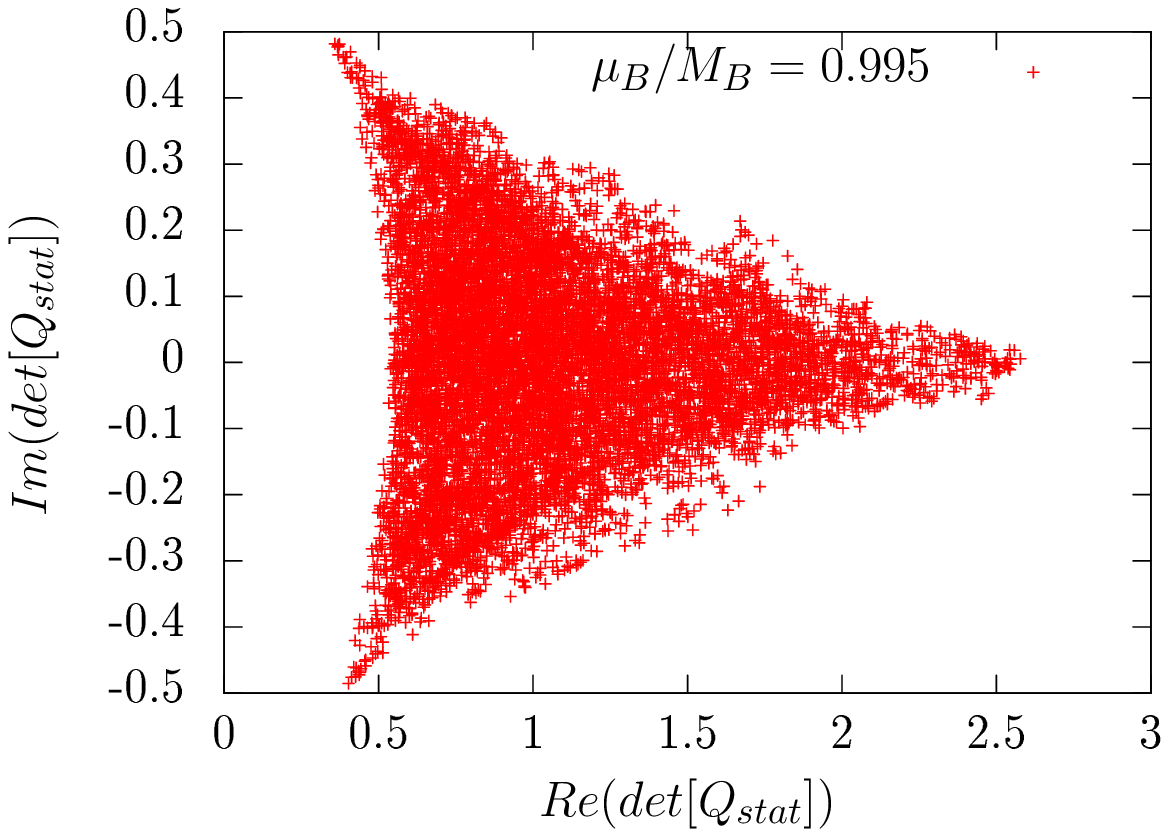}
\includegraphics[width=0.5\textwidth]{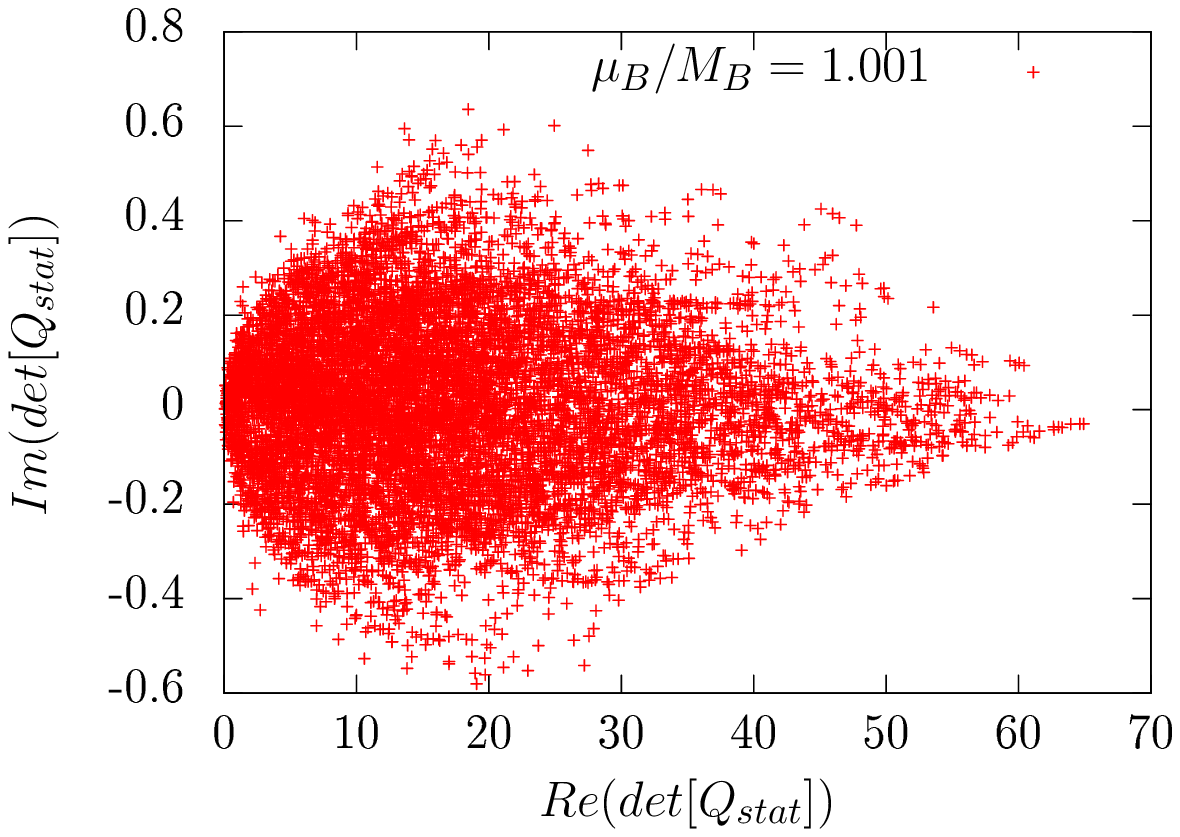}
}
\caption[]{Distribution of the static determinant, eq.~(\ref{eq_qsim}), in the course of simulations 
with $N_f=1, \kappa=0.0173, N_\tau=100, \beta=0$. No crossings of the negative real axis are observed.}
\label{fig:scatter}
\end{figure}

Another problem related to the distribution in the complexified field space has recently been pointed
out for all partition functions containing a complex determinant \cite{kim13}. Its contribution
to the effective action is $\sim \log \det$, and the logarithm develops a cut along the negative 
real axis, i.e.~it is multi-valued. This may cause a problem whenever the calculation of the 
drift term for the Langevin time requires a derivative to be taken across the cut. In \cite{kim13} it
was found for a random matrix model that these crossings lead to incorrect predictions for 
observables if they happen frequently in a Monte Carlo history. Here we can see another benefit 
of the effective theory compared to a Langevin simulation of full QCD. In the effective theory, only
the static determinant features this problem, while the corrections to the effective action 
in the hopping expansion are exponentials of polynomials. We have 
monitored the static determinant during the Langevin evolution, an example is shown in 
figure \ref{fig:scatter} at baryon density slightly below (left) and above (right) the onset transition to 
nuclear matter. Note that the static determinant is dominated by the Polyakov loop.
One observes the expected distortion from the centre symmetric distribution of the 
vacuum state to the distribution preferring the real centre sector, and this distortion is amplified 
exponentially with chemical potential. For simulation purposes, the crucial observation is that
there are nearly no crossings of the negative real axis, in accord with the satisfied 
convergence criterion above. We have monitored such scatter plots over a wide range of
parameter values. Occasionally crossings of the negative axis do occur, but the observed
frequency was $<10^{-4}$ in all cases.  

\subsection{Comparison with Monte Carlo}

\begin{figure}[t]
\centerline{
\includegraphics[width=0.5\textwidth]{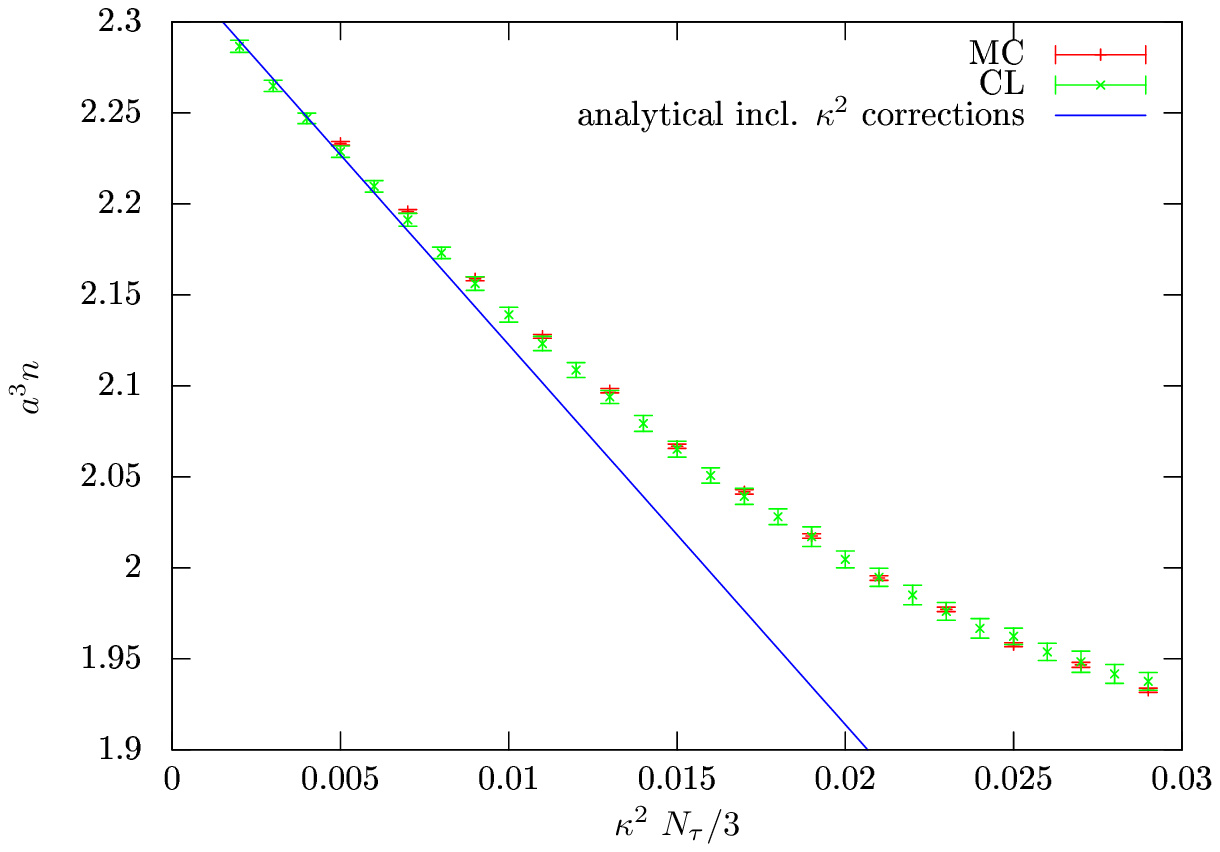}
\includegraphics[width=0.5\textwidth]{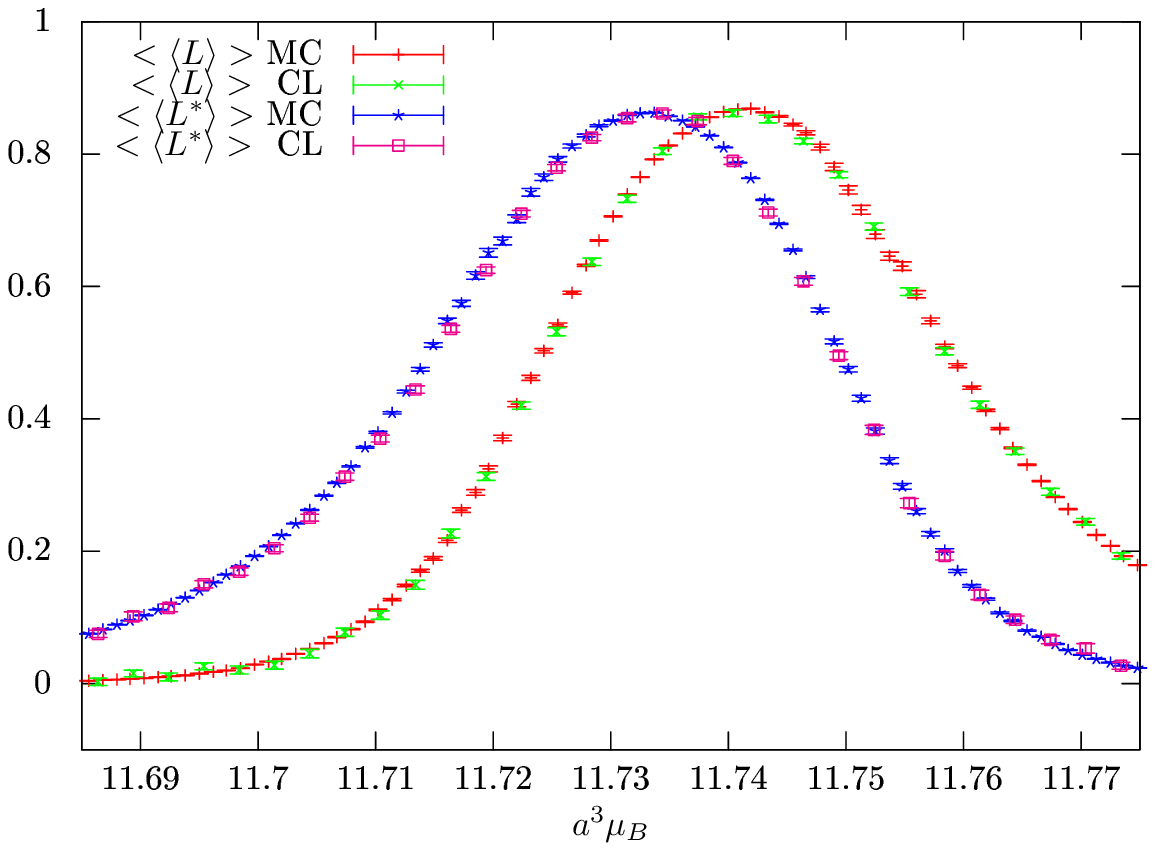}
}
\caption[]{Comparison between Langevin and Monte Carlo for quark number density at different values of $\kappa$ with $N_{\tau} = 100$ and $\beta = 0$ (left)
and the Polyakov loop at different $\mu$ with $\beta=5.7, \kappa=0.01$ and $N_{\tau}=200$ (right), both using the $\kappa^4$-action for $N_f=1$.}
\label{fig:cfMC}
\end{figure}

As a final and complementary check of the validity of the complex Langevin
simulations, one may also  compare with reweighted Monte Carlo results where this is possible, 
i.e.~on small volumes. In \cite{Fromm:2012eb} we have shown a successful comparison 
for very small hopping parameters $\kappa\sim 10^{-4}$. 
Figure \ref{fig:cfMC} shows that this test is also passed 
for significantly larger values $\kappa\sim 0.01$.
We conclude that complex Langevin simulations of the effective theory constructed here
are fully controlled for the entire coupling range investigated, $0<\beta<6$ and $0<\kappa <0.12$.
This is an algorithmic advantage over Langevin simulations in full QCD, where gauge cooling 
techniques \cite{cool} are required to control the field distribution 
and even then simulations at small lattice couplings are ruled out \cite{denes}.

\section{Numerical Results \label{sec:phys}}

\begin{figure}[t]
\centerline{
\includegraphics[width=0.5\textwidth]{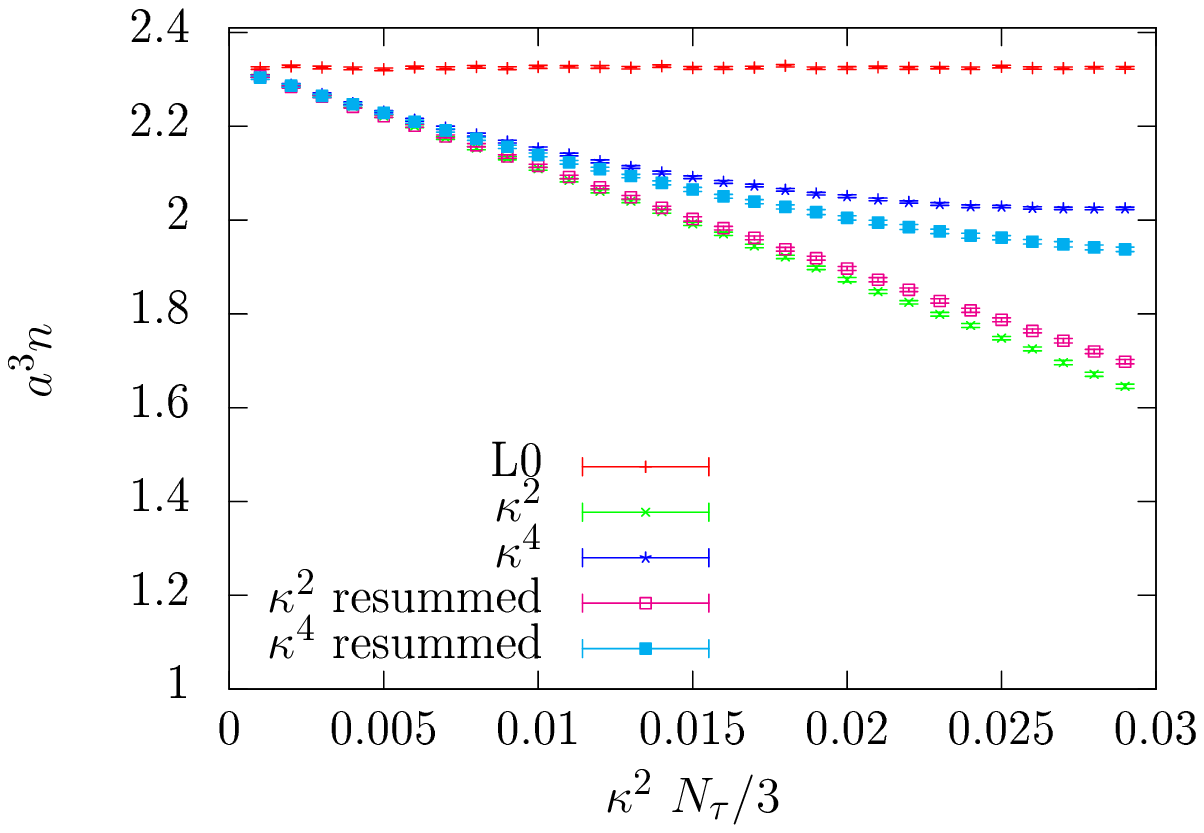}
\includegraphics[width=0.5\textwidth]{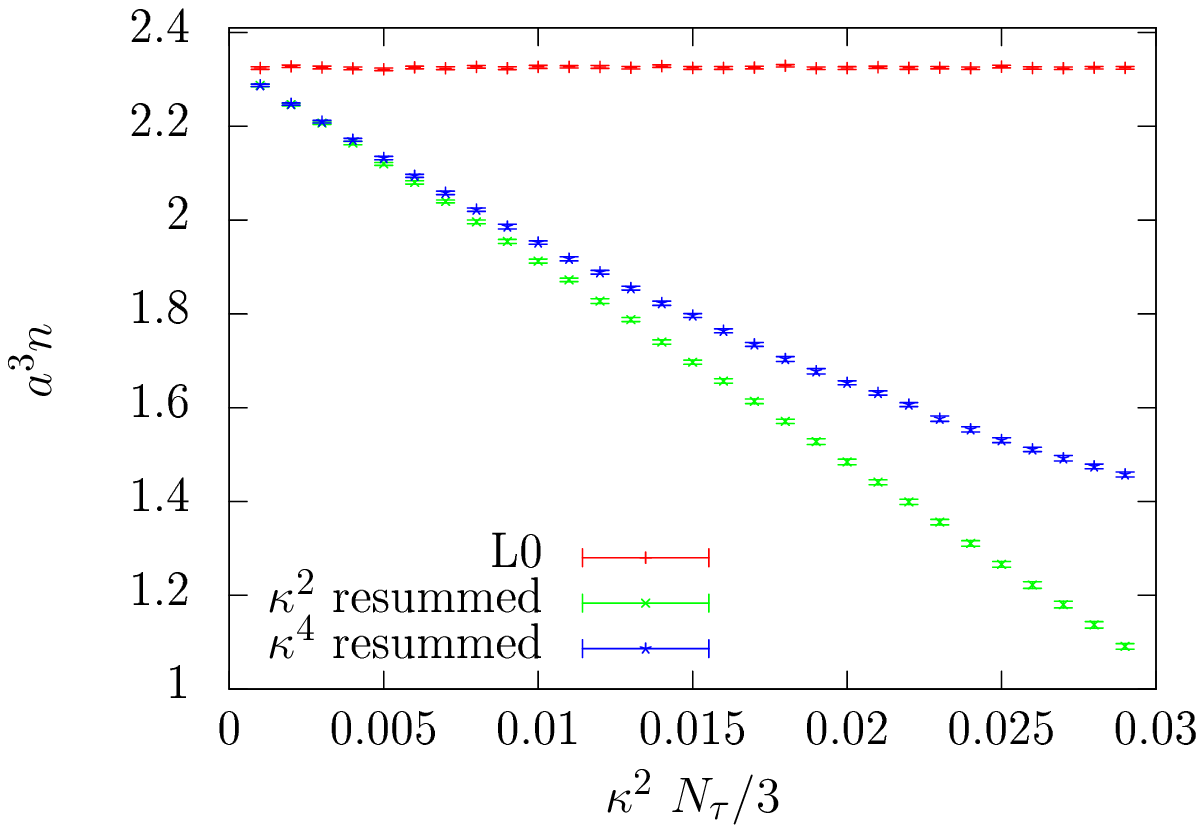}
}
\caption[]{Comparison between $\kappa^2,\kappa^4$ actions, with and without
resummation for $N_f=1,c=0.8$ and $\beta=0$ (left) and resummed, including gauge corrections
for $\beta=5.7$ (right).}
\label{fig:convergence}
\end{figure}

\subsection{Convergence region of the hopping series}

An important task is to find the region of validity
of the effective theory. By this we mean the region, determined by a self-consistent test, 
where the truncated effective theory is a 
good approximation to the full theory.
As criteria we choose the difference between expectation values of observables,
calculated from the $\kappa^2$ and the $\kappa^4$ action, 
$\langle O\rangle_{\kappa^2}, \langle O \rangle_{\kappa^4}$. These can 
be evaluated as a function of the expansion
parameter $\frac{\kappa^2 N_{\tau}}{N_c}$, and the convergence region is where the difference
is smaller than the desired accuracy.
Since we are interested in the onset of baryon number,
we choose the density in lattice units $a^3 n$ as an observable and compute it at a fixed
value of the coupling $h_1 = 0.8$.
As can be seen in figure \ref{fig:convergence}, the static limit is only a valid
approximation in 
the $\kappa \to 0$ limit.  Note
that the resummed 
action offers a slightly better convergence. Therefore, we will use this version for our
simulations.
The expansion parameter already shows that the region of 
convergence is limited in the direction of low temperatures and light quarks,
i.e.~one can reach lower quark masses at larger temperatures.

\subsection{Setting a scale and units}

Setting a scale and performing continuum limits along lines of constant physics
is a computationally very demanding task. Rigorously speaking, this is truly possible only at 
or near the physical point. On the other hand, the 
effective theory considered here is only valid for very 
heavy quarks, due to the truncated hopping series. While it exhibits most qualitative features of
physical QCD, its spectrum is still far from the experimentally 
observed one. For this reason we do not attempt to accurately fix our hadron masses. (In the
mass ranges considered this would anyway demand heavy quark effective theories \cite{hqet}). 
Instead we only provide
a very rough guide where we are in parameter space. 

Our procedure is as follows:
heavy quarks have little influence on the running of the coupling. Thus we use the non-perturbative beta-function of pure gauge theory 
for the lattice spacing in units of the Sommer parameter, $a(\beta)/r_0$ \cite{sommer}.
Taking $r_0=0.5~{\rm fm}$ sets a physical scale for our lattices, while $N_\tau$ tunes
temperature via $T=(aN_\tau)^{-1}$. In a {\it very rough} approximation we then use the
strong coupling expressions eq.~(\ref{eq:hadron}) for the hadron masses.  

\subsection{The nuclear liquid gas transition in heavy dense QCD}

\begin{figure}[t]
\centerline{
\includegraphics[width=0.5\textwidth]{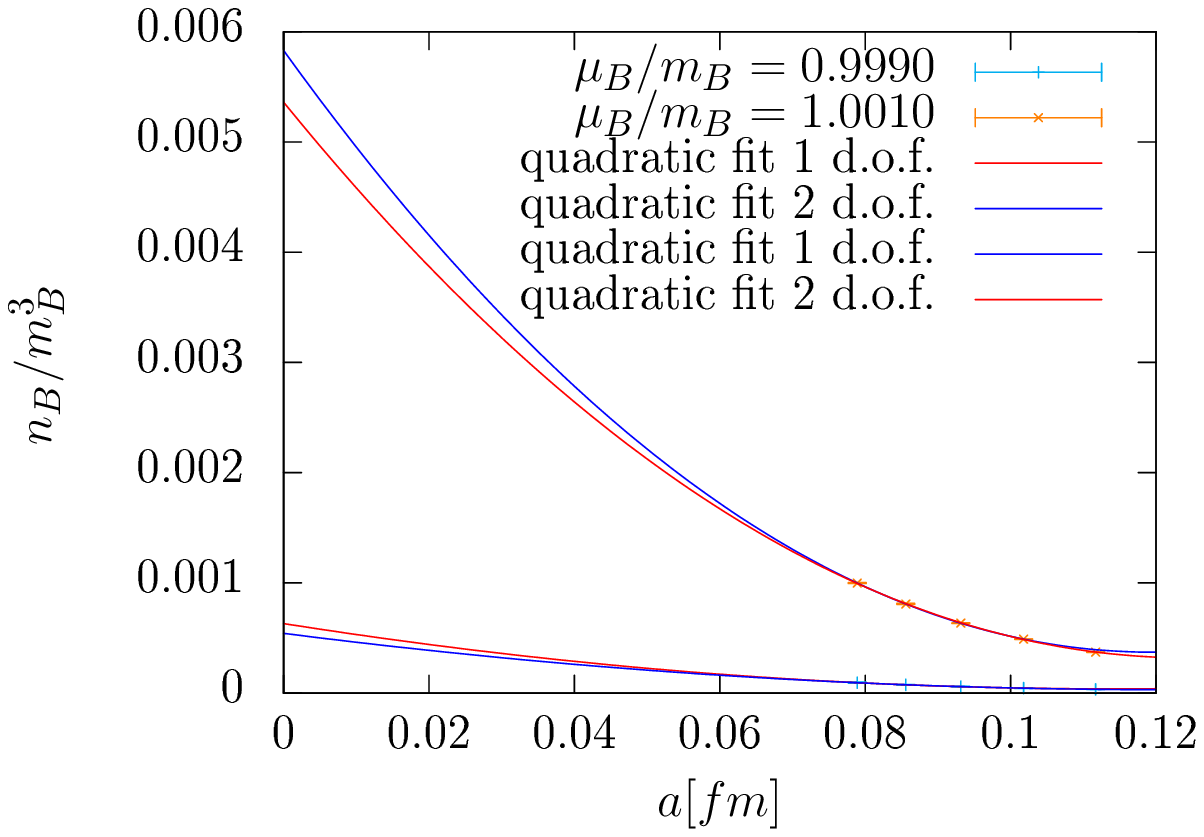}
\includegraphics[width=0.5\textwidth]{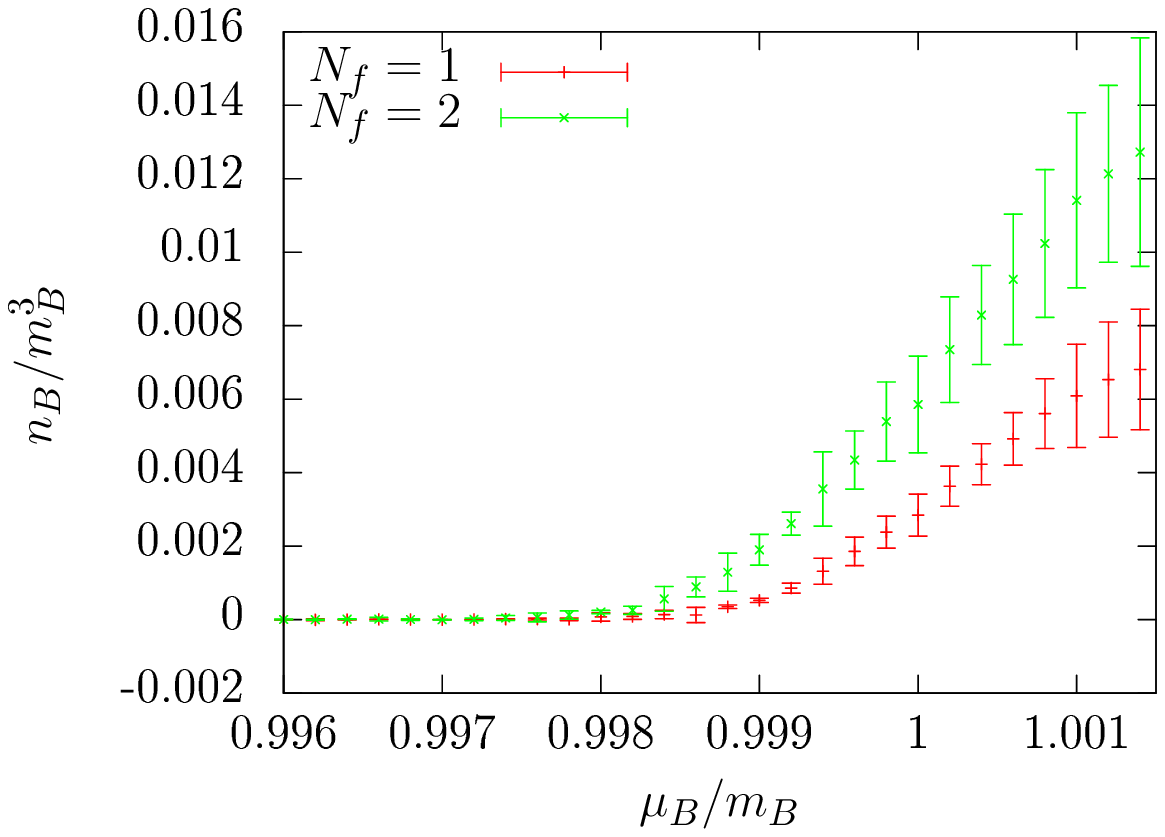}
}
\caption[]{
Example for the continuum extrapolation for $N_f=2$ (left). 
Shown are extrapolations with one d.o.f.  
Continuum extrapolated results for the transition to cold nuclear matter
for T=10MeV and one or two flavours (right). }
\label{fig:silver}
\end{figure}
\begin{figure}[t]
\centerline{
\includegraphics[width=0.5\textwidth]{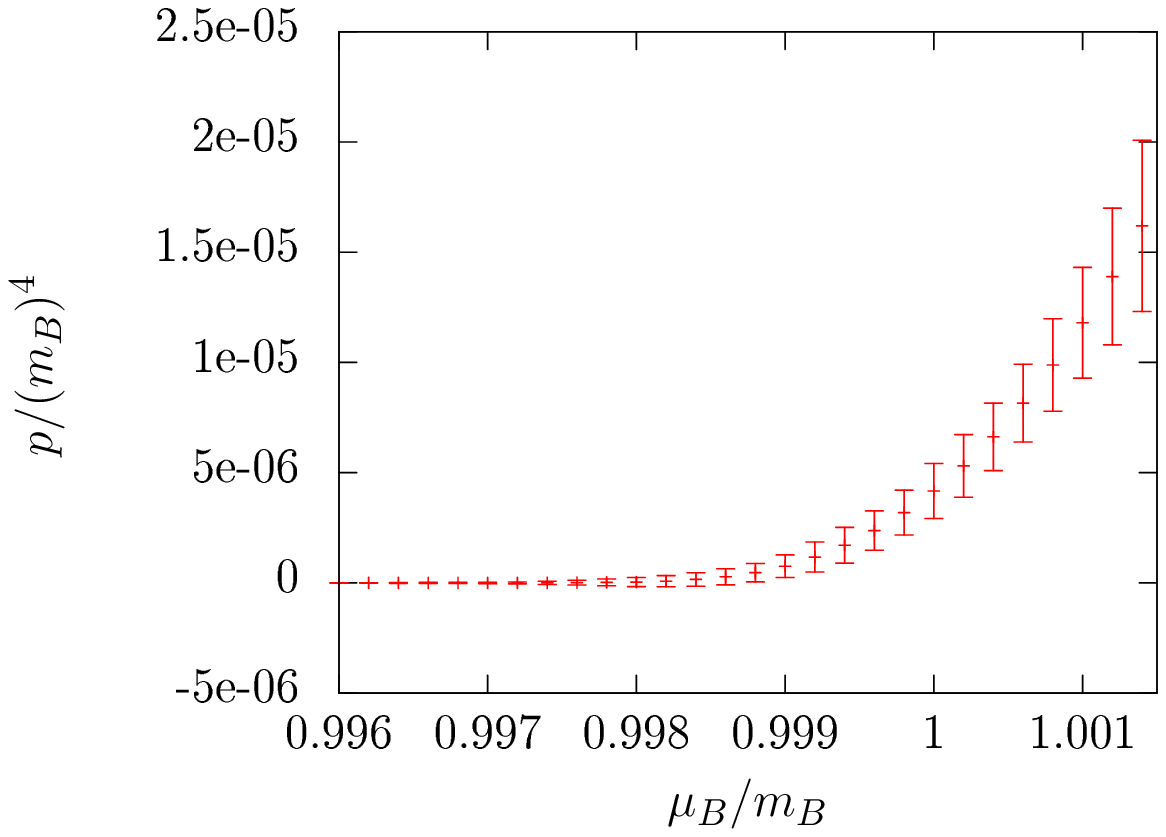}
\includegraphics[width=0.5\textwidth]{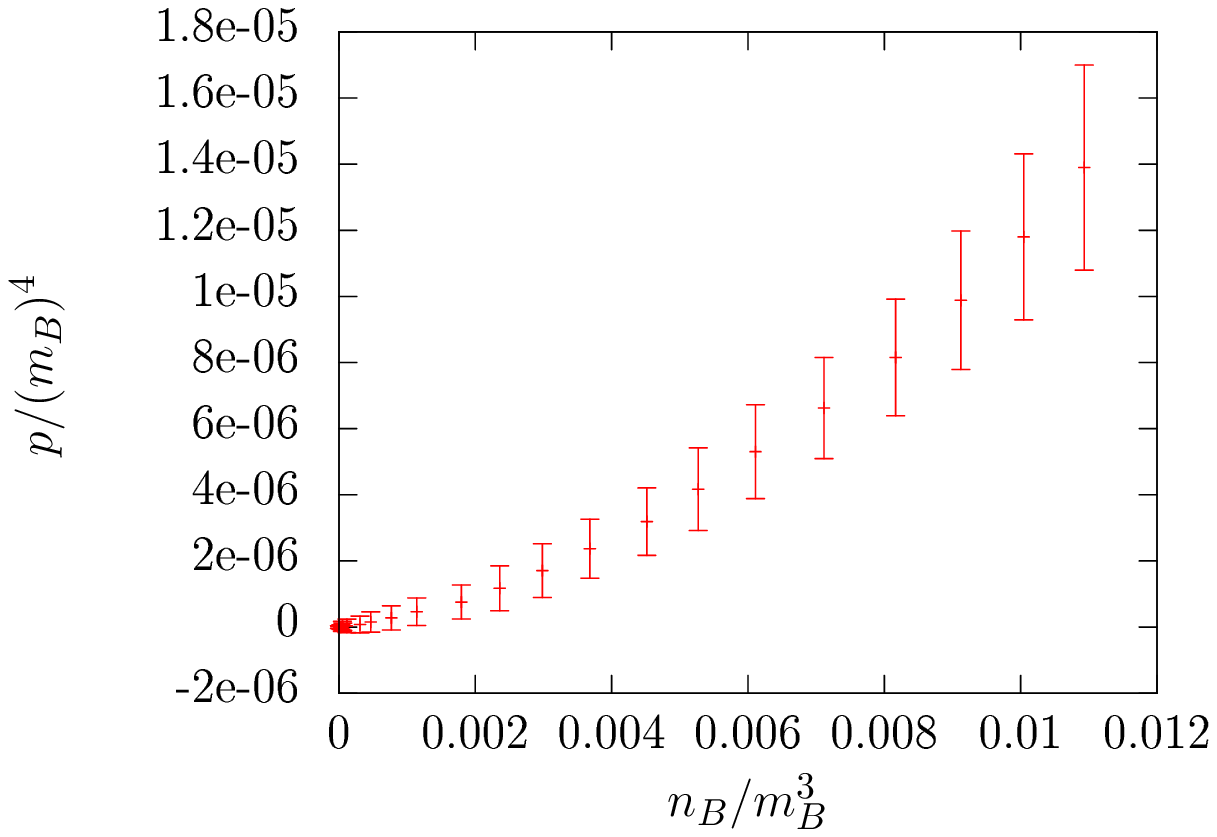}
}
\caption[]{
Pressure and equation of state for $N_f=2$ at $T=10$ MeV.}
\label{fig:eos}
\end{figure}

In our previous work \cite{Fromm:2012eb} we performed a continuum extrapolation for
the transition 
to cold nuclear matter based on the $\kappa^2$ action. 
In figure \ref{fig:convergence} we repeat this calculations including the 
$\kappa^4$ corrections. This allows us to simulate smaller lattice spacings 
$a=0.08$ fm without leaving the 
region of convergence, since reducing $a$ while keeping $m_B/T$ and $T$ fixed means
going to higher 
$\kappa$ and $N_{\tau}$. Nevertheless, the 
extrapolation suffers from considerable uncertainties, resulting in large errors in
the high density phase. 
This can be seen in fig. \ref{fig:silver} (left), where we show the two best fits for our data at 
$\mu_B/m_B=1$ at several lattice spacings. 
This is the chemical potential where different extrapolation fits differ the most.
The systematic truncation error for our $\kappa^4$ data is estimated as the difference to the data obtained from the $\kappa^2$ action and included in the error bars in the figure. This data was  then fitted to get a value for $a \rightarrow 0$. For each value of the chemical potential we tried several fits (linear and quadratic) with one to three degrees of freedom. For the best fits we always achieved 
$\chi^2_{red}<2$ as long as $\mu_B/m_B < 1.0014$.
For the continuum result we quote the average of the two best fits, the error was estimated as difference between those two fits.
We note that the results at $\kappa^4$ are somewhat higher than  
our $\kappa^2$-results in \cite{Fromm:2012eb}. This is because inclusion of $\kappa^4$ is the first
order allowing for a realistic estimate of
the truncation error, and thus permits inclusion of data with 
smaller lattice spacing. 

This results in the continuum extrapolated baryon number density in figure
\ref{fig:silver} (right), where we display the results for $N_f=1,2$ for a temperature $T=10$ MeV. 
In the low density region the "silver blaze" property, i.e.~the independence of the thermodynamic 
functions of chemical potential can be seen. 
The growing uncertainties in the high density region are caused by the unphysical 
saturation on the lattice which limits the density to $2 N_f N_c$ quarks per 
lattice site, while in the continuum no such saturation exists.
As expected, the onset of nuclear matter happens at a critical value $\mu_B^c<m_B$, 
due to the nuclear binding energy. The location of the onset suggests a very small binding energy
$\sim 10^{-3} m_B$ for the heavy quarks considered here, 
in accord with our perturbative analysis, section \ref{sec:pt}. This explains why the 
onset transition is a smooth crossover rather than the first-order transition expected for light quarks.
The endpoint of the nuclear liquid gas transition sits at a temperature of the order of the binding
energy and is not visible for very heavy quarks. In accord with expectation, the onset with two flavours
is steeper than with one flavour. 

\begin{figure}[t]
\centerline{
\includegraphics[width=0.5\textwidth]{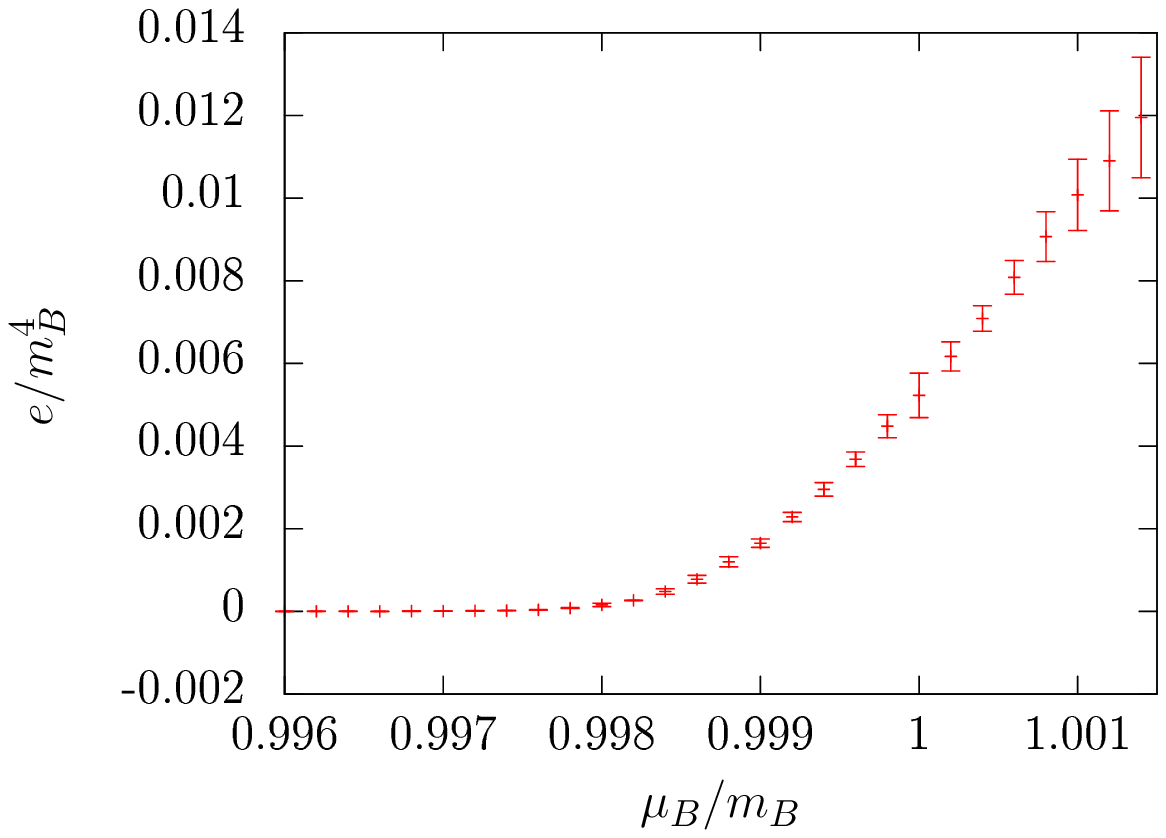}
\includegraphics[width=0.5\textwidth]{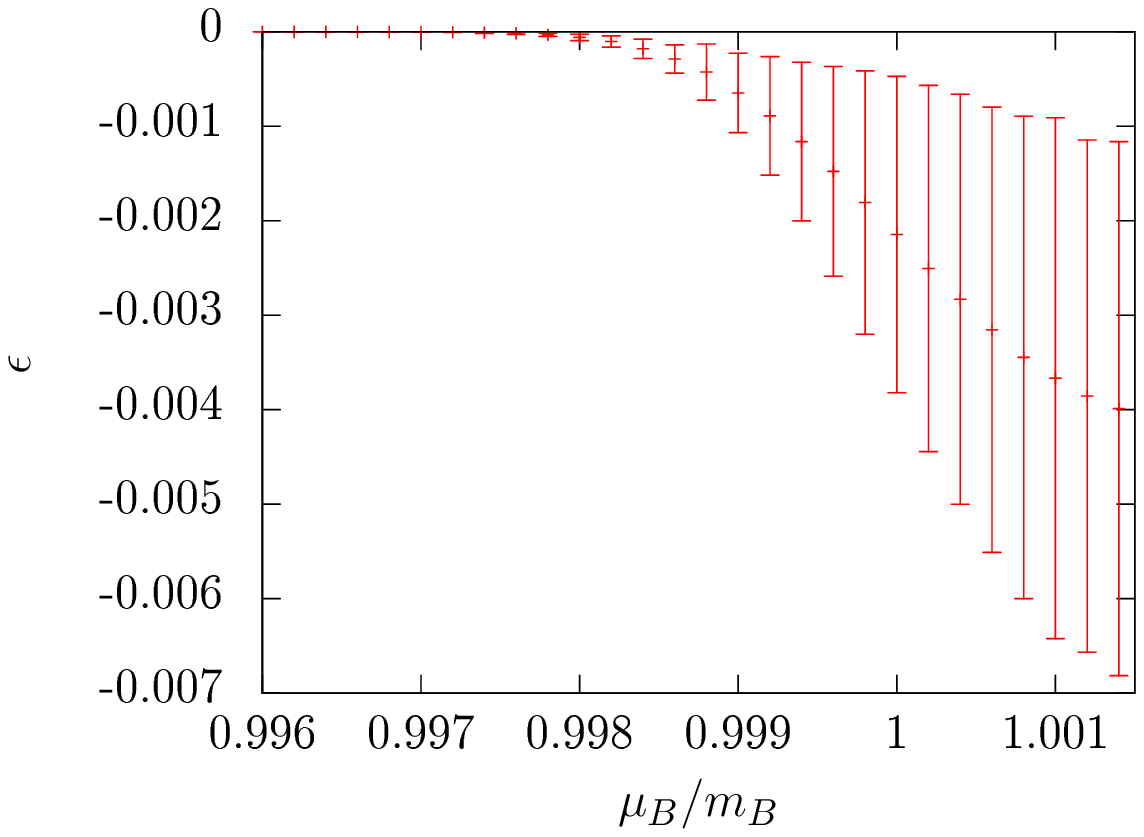}
}
\caption[]{Left: Energy density, eq.~(\ref{eq:e-density}). Right: Binding energy per nucleon, eq.~(\ref{eq:bind}). Both plots show $N_f=2, T=10$ MeV. }
\label{fig:ebind}
\end{figure}
It is now straightforward to compute the other thermodynamic functions and from them the equation
of state. Figure \ref{fig:eos} shows the pressure as a function of baryon chemical potential as well
as a function of baryon density, whereas the binding energy per nucleon is shown in 
figure \ref{fig:ebind}. Note that in all plots the error bars include the systematic uncertainty of both,
the truncation of the effective theory as well as the continuum extrapolation. The plot of the
binding energy is particularly intriguing. For small density it is zero, another manifestation of the silver 
blaze property, until it turns negative, thus causing the condensation of nuclear matter. 
At larger density, lattice saturation is reached before the expected
upturn of the curve. Nevertheless, the shape of the curve suggests that the minimum has been 
reached near the right border. Its numerical value of the order of $10^{-3}$ is 
consistent with that observed from the location of the onset transition in figure \ref{fig:silver} (right).

\subsection{Nuclear liquid gas transition for light quarks}

As in our previous work \cite{Fromm:2012eb}, the accessible quark masses in the convergence region of the effective theory are 
too high to realise the expected first order transition for the onset of nuclear matter. 
Finite size scaling
analyses reveal the transition to be a smooth crossover, in accord with the interplay between
accessible temperatures and the values of the binding energies.
Of course it is highly interesting to see whether the effective theory includes the expected  
physics features when the quark mass is lowered. 
We now consider $\kappa = 0.12$, corresponding to a small quark mass, and very 
low temperatures parametrised by $N_{\tau} \sim O(10^3)$. We stress that this choice of parameters 
is far outside the convergence region of our $\kappa^4$-action, cf.~figure \ref{fig:convergence}. 
In other words, there is no reason to expect the results to accurately represent QCD and 
an attempt at a continuum extrapolation makes no sense. Nevertheless, this is an interesting
check of the qualitative features of the effective theory. 

\begin{figure}[t]
\centerline{
\includegraphics[width=0.35\textwidth]{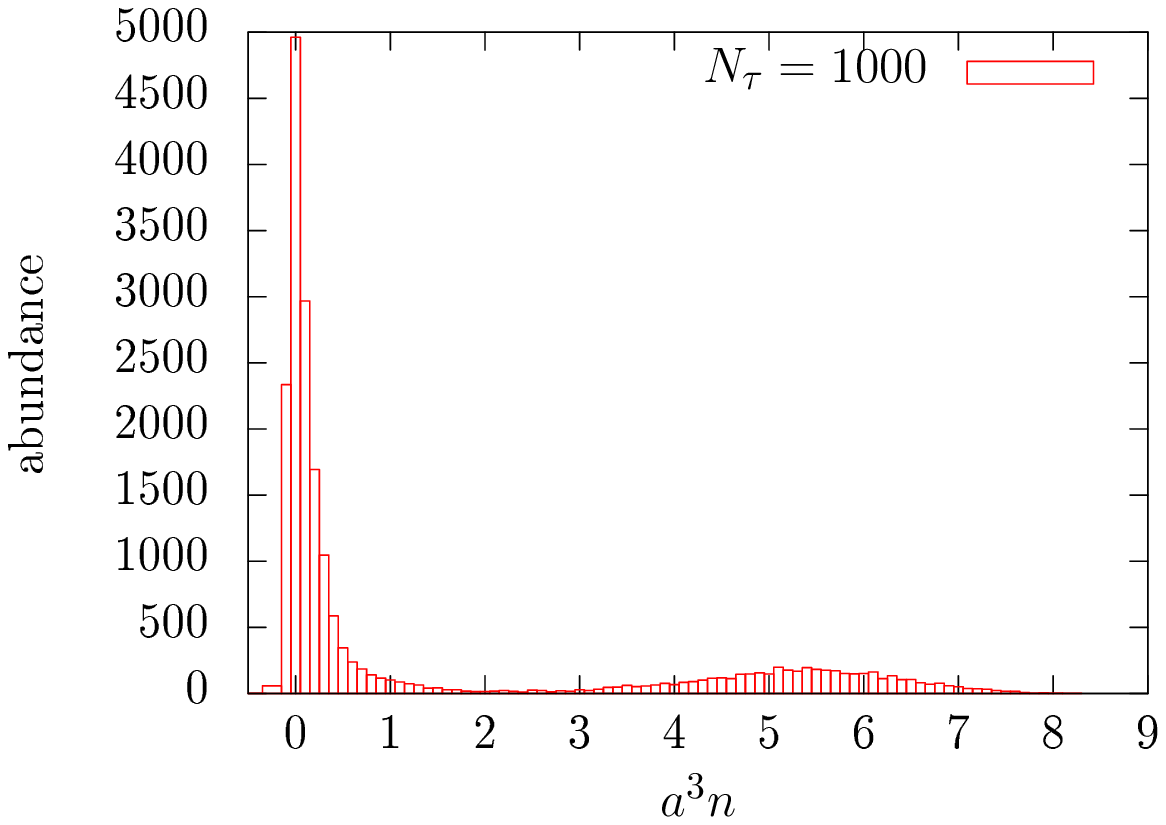}
\includegraphics[width=0.35\textwidth]{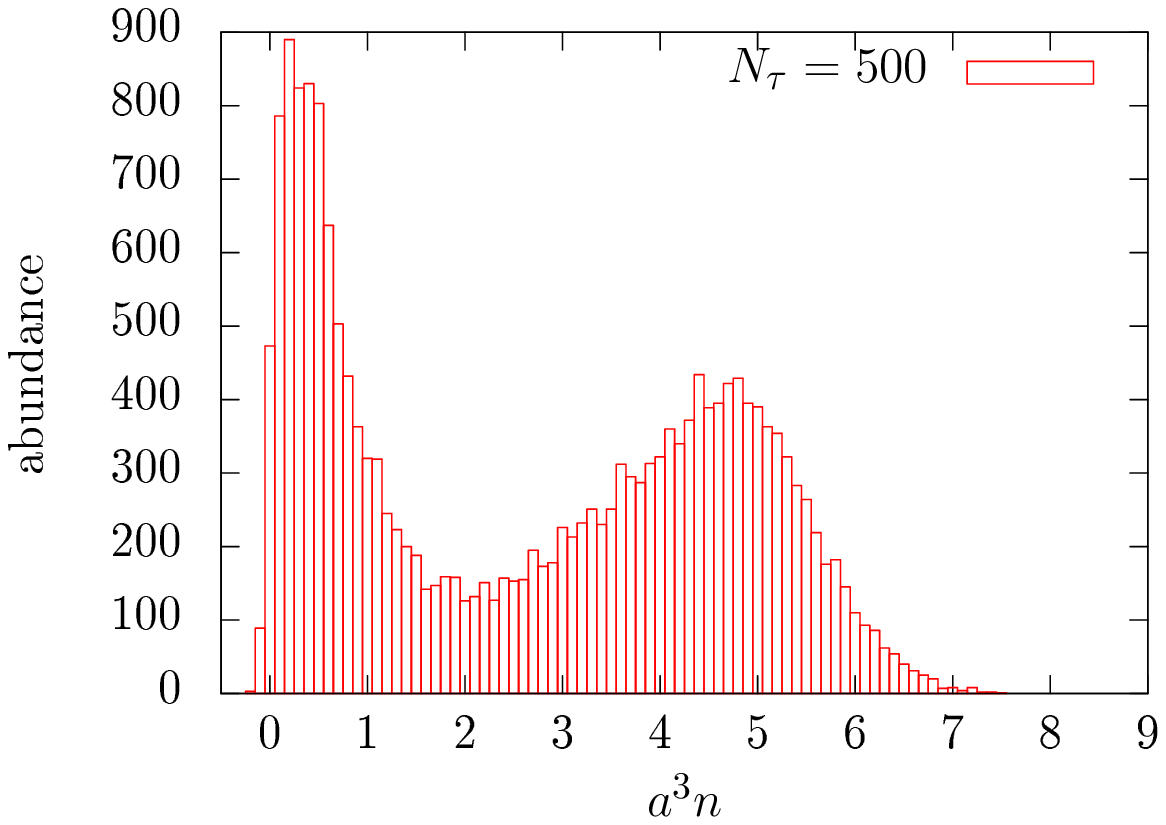}
\includegraphics[width=0.35\textwidth]{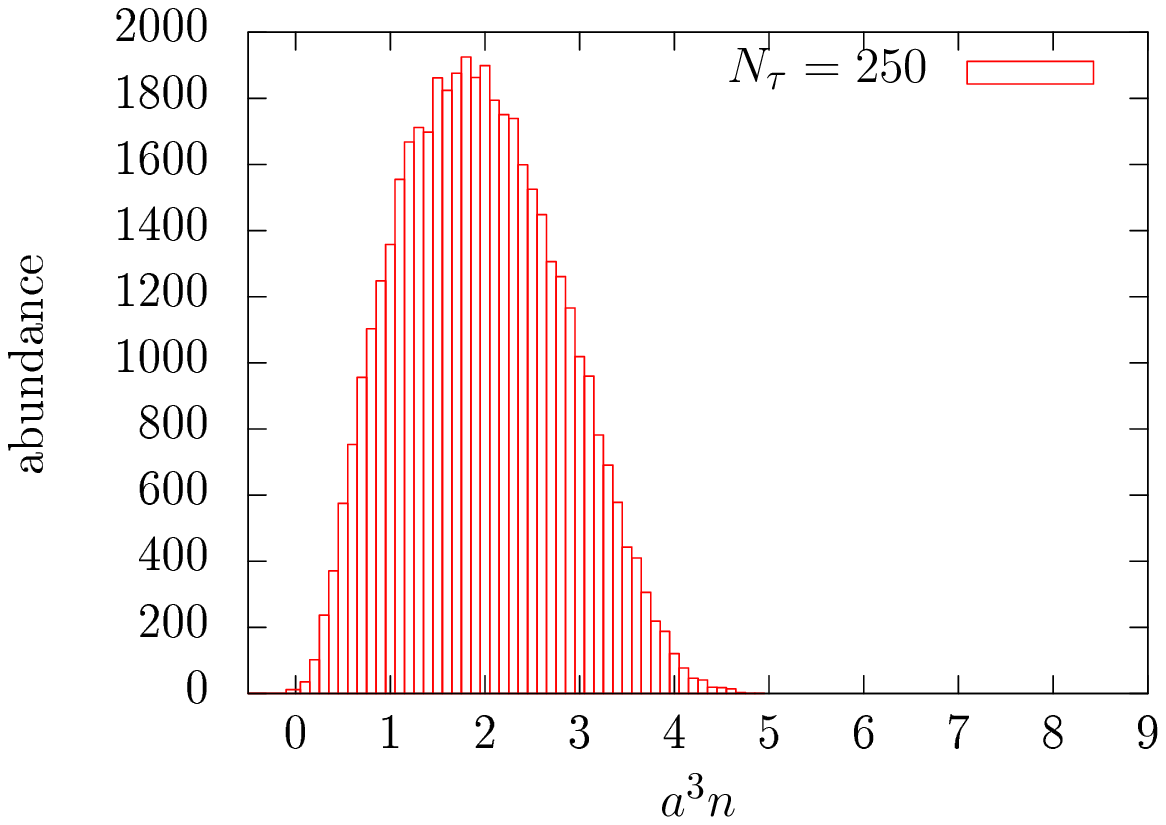}
}
\caption[]{Distributions of the quark density in the transition region with temperature increasing from left to right, 
$\kappa = 0.12$ and $\beta = 5.7$ }
\label{fig:polyakov-hist}
\end{figure}
\begin{figure}[t]
\centerline{
\includegraphics[width=0.5\textwidth]{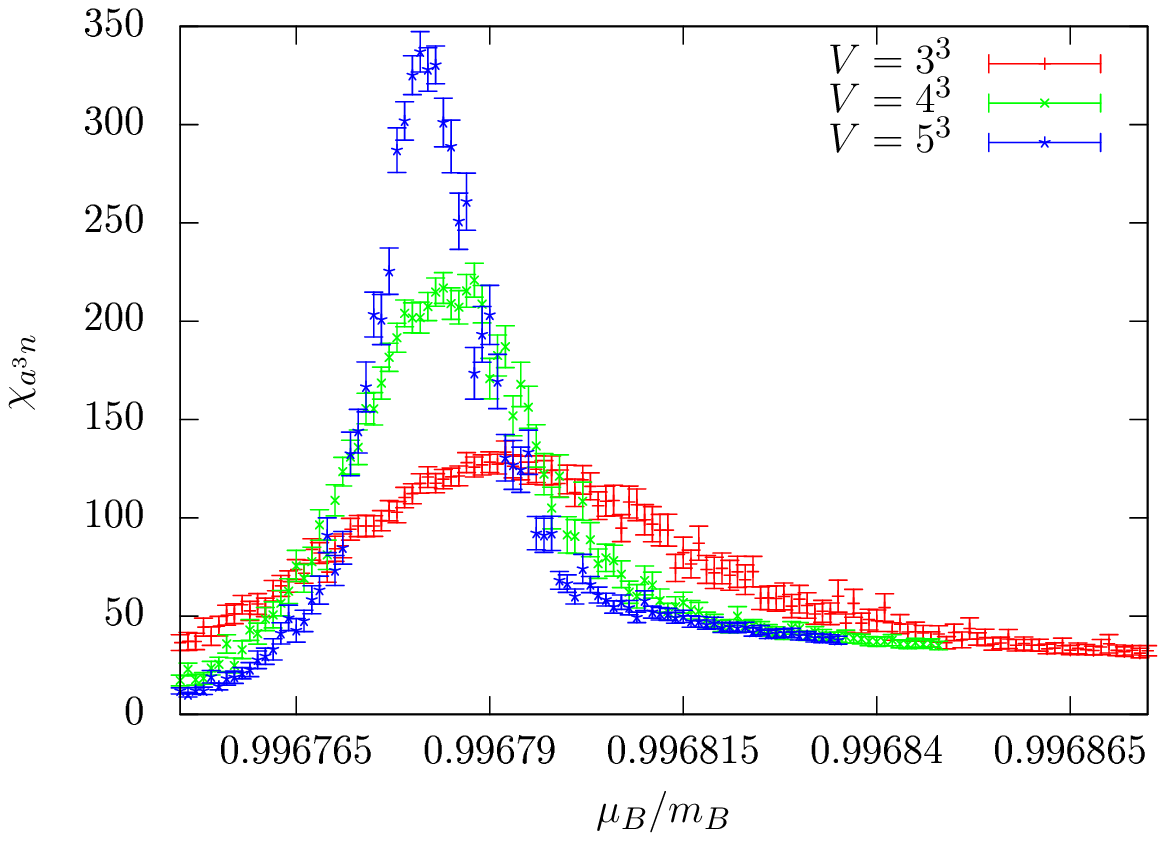}
\includegraphics[width=0.5\textwidth]{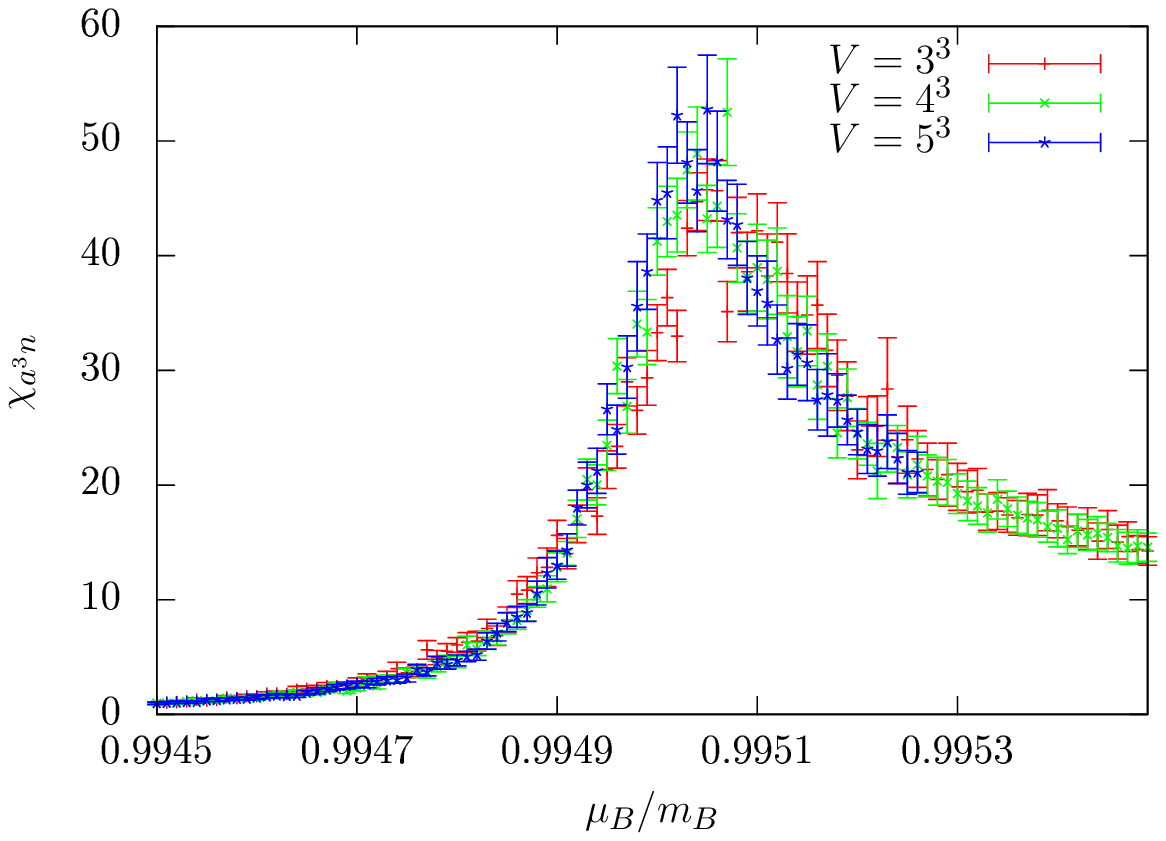}
}
\caption[]{Quark number susceptibility for $\kappa = 0.12$ and $\beta = 5.7$ and
 $N_{\tau} = 500$ (left) and $N_\tau=250$. The divergence with volume signals a true phase
 transition, whereas saturation at a finite value implies a smooth crossover.}
\label{fig:polyakov-susc}
\end{figure}

Figure \ref{fig:polyakov-hist} shows distributions of the Polyakov loop
in the onset transition region for three choices of $N_\tau$, corresponding to increasing temperatures
from left to right. We clearly observe the coexistence of two phases at the lowest temperatures, which
indicates a first order transition between them.  
As the temperature is raised ($N_\tau$ is lowered), the two-state signal weakens and merges to a
single gaussian distribution, signalling a weakening and eventual disappearance of the first-order
transition. This picture is corroborated by a finite size analysis of the quark number susceptibility in 
figure \ref{fig:polyakov-susc}. First-order and crossover transition are clearly distinguished by diverging
and finite susceptibility  as  a function of volume. Thus we conclude, while our $\kappa^4$-action 
used in this work is not quantitatively reliable in this parameter range, it displays all the qualitative
features expected for the nuclear liquid gas transition: a first-order transition from the vacuum to
nuclear matter which weakens with temperature until it vanishes in a critical endpoint. We therefore 
expect higher orders in the effective action to only correct the quantitative details of this transition.

\subsection{Isospin vs.~baryon chemical potential}

\begin{figure}[t]
\centerline{
\includegraphics[width=0.5\textwidth]{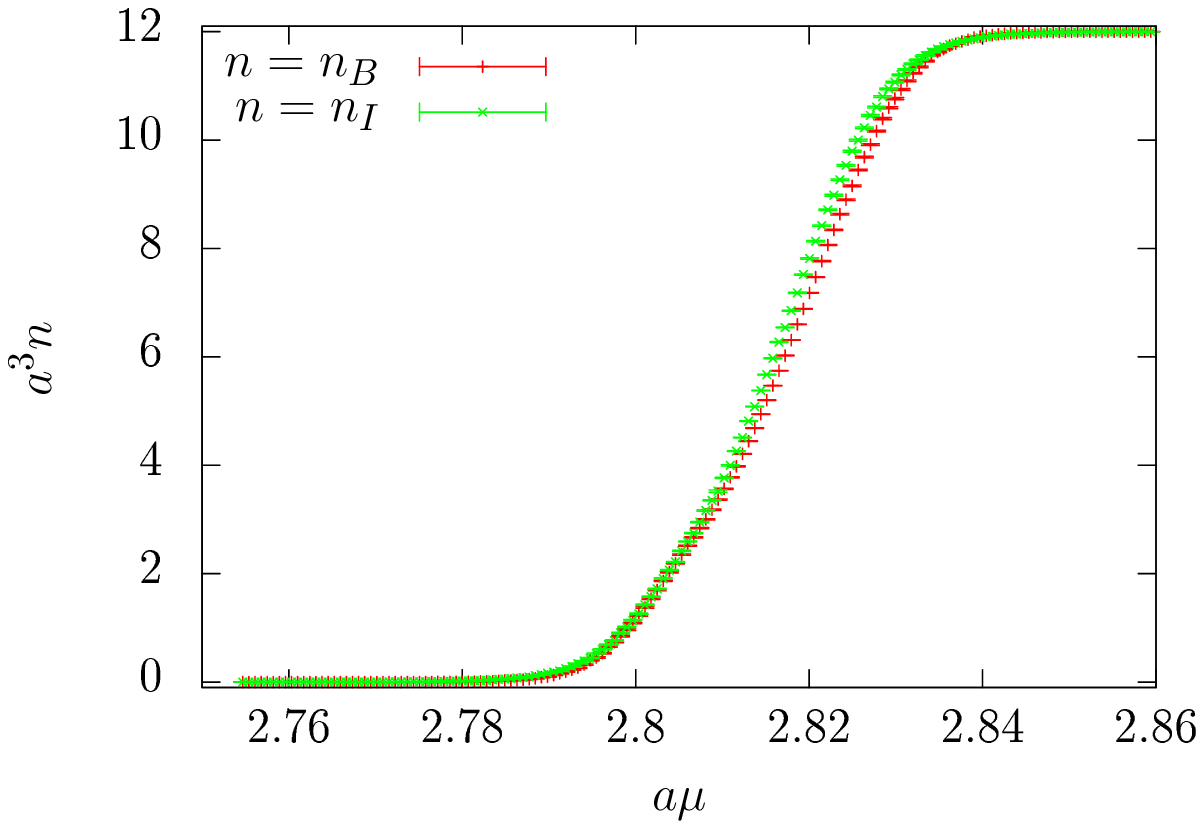}
\includegraphics[width=0.5\textwidth]{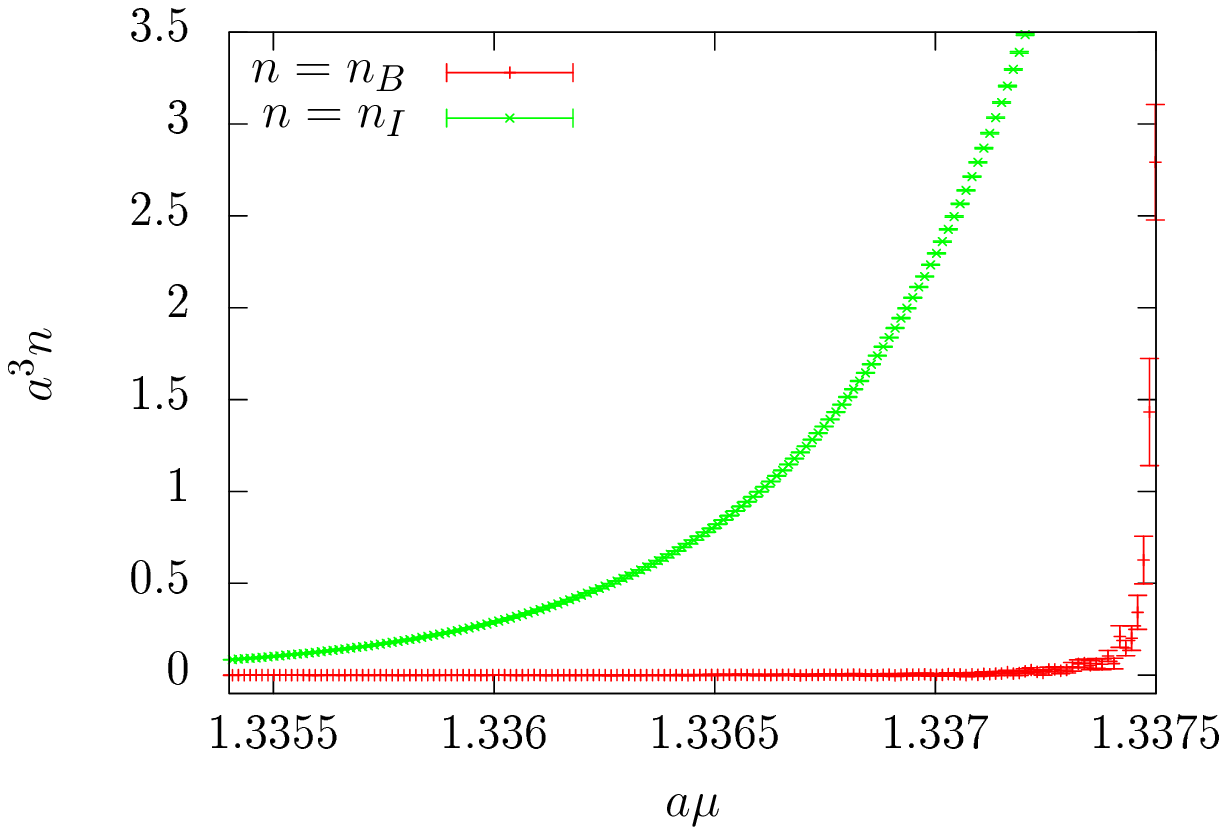}
}
\caption[]{Onset of finite isospin density vs. baryon density for $N_f=2, N_\tau=100, \beta=5.7$ and
heavy quarks, $\kappa=0.03$ (left) and light quarks, $\kappa=0.12$ (right).
}
\label{fig:iso-bar}
\end{figure}
Let us finally consider the situation in the two-flavour theory 
with finite isospin chemical potential, $\mu_I=\mu_u=-\mu_d$. In section \ref{sec:iso} we have 
discussed the situation in the static strong coupling limit, where the onset transition for 
pion condensation at $\mu_I=m_\pi/2$ happens at the same chemical potential as 
the one for baryon condensation at $\mu_B=\mu_B/3$.  With interactions included, this gets modified
in two ways. Firstly, we have $m_\pi/2 < m_B/3$ in this case, and secondly the onset gets shifted to
smaller chemical potentials by the non-vanishing binding energy. The first effect also leads to 
the expected gap opening between the onset to pion condensation vs.~that to baryon condensation \cite{cohen}, when plotted
against quark chemical potential, as shown in figure \ref{fig:iso-bar}. 

\section{Conclusions}

In this work we further elaborated the construction of an effective three-dimensional lattice theory
for QCD thermodynamics.
It is formulated entirely in terms of Polyakov loops and calculated from the 4d Wilson action as a strong coupling and hopping series
which is now complete to order $\kappa^nu^m, (n+m)=4$. In the static strong coupling limit, the 
effective theory can be solved exactly, providing the complete spin-flavour structure of the hadron 
spectrum as well as an onset transition from zero density to lattice saturation.
The interacting  
effective theory has a sign problem that can be handled by complex Langevin simulations with fully
satisfied convergence criteria. Moreover, the sign problem is mild enough that on small volumes
Monte Carlo simulations are feasible, even at real chemical potential. The couplings of the effective theory
are sufficiently small to also permit a perturbative evaluation, which agrees with numerical results
in wide regions of the parameter space. Altogether this allows for a controlled and very efficient
evaluation of thermodynamic functions and critical couplings.

Working in the heavy quark region near the static limit, where  
continuum extrapolations of thermodynamic functions are feasible, we have explicitly demonstrated
the onset transition to cold nuclear matter 
and calculated the nuclear equation of state for the first time directly from QCD. In particular, we 
find a negative binding energy per nucleon as the expected reason for baryon condensation. In accord
with expectations from models of nuclear interactions, the binding energy is governed by exponentials
of the meson mass and suppressed for heavy quarks. Decreasing the quark mass beyond the convergence 
region of our expansion, we indeed observe the nuclear onset transition to emerge as a first order 
liquid gas transition with an endpoint at some small temperature. In this parameter range also the expected
gap opens up between the onset of pion condensation in the case of finite isospin chemical potential
and the nuclear onset at finite baryon density.

In summary, the effective lattice theory described in this work contains all the qualitative physics 
expected for cold nuclear matter. 
It remains to be seen whether high enough orders of the hopping 
expansion can be generated in the future in order to reach physical quark mass values. 
However, since the hopping
convergence is much faster at high temperatures, the current effective theory might already be
useful to describe the finite temperature phase structure of QCD with light quarks.
Work in this direction is in progress.

\section*{Acknowledgements}
We thank Georg Bergner for providing the Monte Carlo data for figure \ref{fig:cfMC} and are indebted to
Georg Bergner, Jonas Glesaaen and Wolfgang Unger for innumerable discussions, checks, proof 
reading and advice.
J.L. is supported by the Swiss National Science Foundation under 
grant 200020-137920. M.N. and O.P. are  partially supported by the German BMBF, 
grant 06FY7100, and the Helmholtz International
Center for FAIR within the LOEWE program launched by the State of Hesse. 

\appendix
\section{Wilson line contributions to the effective action}

In this appendix we list final expressions for all types of terms appearing in the kinetic determinant 
derived in section \ref{sec:gi}.
\begin{align}
-\frac12 \int [dU_k] &\sum_{\vec{x},i} \text{Tr} P_{\vec{x}, i} M_{\vec{x}, i} P_{\vec{x}, i} M_{\vec{x}, i} = \\
\frac{\kappa^4 N_{\tau} (N_{\tau} - 1)}{N_c^2} 
\sum_{\vec{x},i} 
\Big\{
& \text{Tr} \Big(
 \frac{h_1 W_{\vec{x}}}{(1 + h_1 W_{\vec{x}})^2} 
+ \frac{\bar{h}_1 W^{\dagger}_{\vec{x}}}{(1 + \bar{h}_1 W^{\dagger}_{\vec{x}})^2} 
+ 2\frac{\frac{1}{N_{\tau}-1}\sum_{t=1}^{N_{\tau}-1} (2 \kappa)^{2t}}{(1 + h_1 W_{\vec{x}}) (1 + \bar{h}_1 W^{\dagger}_{\vec{x}})}
\Big) \nn \\
\Big(
& \text{Tr} 
\frac{h_1 W_{\vec{x}+i}}{1 + h_1 W_{\vec{x}+i}} 
-
\text{Tr}
\frac{\bar{h}_1 W^{\dagger}_{\vec{x}+i}}{1 + \bar{h}_1 W^{\dagger}_{\vec{x}+i}} 
\Big)^2 
+\Big(
\text{Tr} 
\frac{h_1 W_{\vec{x}}}{1 + h_1 W_{\vec{x}}} 
-
\text{Tr}
\frac{\bar{h}_1 W^{\dagger}_{\vec{x}}}{1 + \bar{h}_1 W^{\dagger}_{\vec{x}}} 
\Big)^2  \nn \\
& \text{Tr} \Big(
 \frac{h_1 W_{\vec{x}+i}}{(1 + h_1 W_{\vec{x}+i})^2} 
+ \frac{\bar{h}_1 W^{\dagger}_{\vec{x}+i}}{(1 + \bar{h}_1 W^{\dagger}_{\vec{x}+i})^2} 
+ 2 \frac{\frac{1}{N_{\tau}-1} \sum_{t=1}^{N_{\tau}-1} (2 \kappa)^{2t}}{(1 + h_1 W_{\vec{x}+i}) (1 + \bar{h}_1 W^{\dagger}_{\vec{x}+i})}
\Big) 
\Big\} \nn \\
 - \frac{\kappa^4 N_{\tau}}{N_c^2-1} 
\sum_{\vec{x},i}  
 \Big\{
& \text{Tr}
\Big( \frac{h_1 W_{\vec{x}}}{1 + h_1 W_{\vec{x}}} 
- \frac{\bar{h}_1 W^{\dagger}_{\vec{x}}}{1 + \bar{h}_1 W^{\dagger}_{\vec{x}}} \Big)^2
\Big( \text{Tr} \frac{h_1 W_{\vec{x}+i}}{1 + h_1 W_{\vec{x}+i}} 
-
\text{Tr} \frac{\bar{h}_1 W^{\dagger}_{\vec{x}+i}}{1 + \bar{h}_1 W^{\dagger}_{\vec{x}+i}}  \Big)^2
 \nn \\
+ \Big( 
& \text{Tr} \frac{h_1 W_{\vec{x}}}{1 + h_1 W_{\vec{x}}} 
-
\text{Tr} \frac{\bar{h}_1 W^{\dagger}_{\vec{x}}}{1 + \bar{h}_1 W^{\dagger}_{\vec{x}}}  \Big)^2
\text{Tr}
 \Big( \frac{h_1 W_{\vec{x}+i}}{1 + h_1 W_{\vec{x}+i}} 
- \frac{\bar{h}_1 W^{\dagger}_{\vec{x}+i}}{1 + \bar{h}_1 W^{\dagger}_{\vec{x}+i}} \Big)^2
\Big\} \nn \\
 + \frac{\kappa^4 N_{\tau}}{N_c^3-N_c} 
\sum_{\vec{x},i} 
 \Big\{
& \text{Tr}
 \Big( \frac{h_1 W_{\vec{x}}}{1 + h_1 W_{\vec{x}}} 
 - \frac{\bar{h}_1 W^{\dagger}_{\vec{x}}}{1 + \bar{h}_1 W^{\dagger}_{\vec{x}}} \Big)^2
\text{Tr} 
 \Big( \frac{h_1 W_{\vec{x}+i}}{1 + h_1 W_{\vec{x}+i}}
 - \frac{\bar{h}_1 W^{\dagger}_{\vec{x}+i}}{1 + \bar{h}_1 W^{\dagger}_{\vec{x}+i}} \Big)^2 \nn \\
+
\Big( 
&\text{Tr} \frac{h_1 W_{\vec{x}}}{1 + h_1 W_{\vec{x}}} 
- \text{Tr} \frac{\bar{h}_1 W^{\dagger}_{\vec{x}}}{1 + \bar{h}_1 W^{\dagger}_{\vec{x}}}  \Big)^2
\Big( \text{Tr} \frac{h_1 W_{\vec{x}+i}}{1 + h_1 W_{\vec{x}+i}}
-\text{Tr} \frac{\bar{h}_1 W^{\dagger}_{\vec{x}+i}}{1 + \bar{h}_1 W^{\dagger}_{\vec{x}+i}}  \Big)^2 
 \Big \} \nn
\end{align}

\begin{align}
\frac12 \int [dU_k]  \sum_{\vec{x}, i} & \text{Tr} P_{\vec{x}, i} M_{\vec{x}, i}  \text{Tr} P_{\vec{x}, i} M_{\vec{x}, i} = \\
 2 \frac{\kappa^4 N_{\tau} (N_{\tau} -1)}{N_c^2} 
\sum_{\vec{x}, i} 
\Big\{
\Big(
& \text{Tr} 
\frac{h_1 W_{\vec{x}}}{1 + h_1 W_{\vec{x}}} 
- \text{Tr}
\frac{\bar{h}_1 W^{\dagger}_{\vec{x}}}{1 + \bar{h}_1 W^{\dagger}_{\vec{x}}} 
\Big)^2 
\Big(
\text{Tr} 
\frac{h_1 W_{\vec{x}+i}}{1 + h_1 W_{\vec{x}+i}} 
- \text{Tr}
\frac{\bar{h}_1 W^{\dagger}_{\vec{x}+i}}{1 + \bar{h}_1 W^{\dagger}_{\vec{x}+i}} 
\Big)^2 \nn \\
+ 
&\text{Tr} \Big(
 \frac{h_1 W_{\vec{x}}}{(1 + h_1 W_{\vec{x}})^2} 
+ \frac{\bar{h}_1 W^{\dagger}_{\vec{x}}}{(1 + \bar{h}_1 W^{\dagger}_{\vec{x}})^2} 
\Big)
\text{Tr} \Big(
 \frac{h_1 W_{\vec{x}+i}}{(1 + h_1 W_{\vec{x}+i})^2} 
+ \frac{\bar{h}_1 W^{\dagger}_{\vec{x}+i}}{(1 + \bar{h}_1 W^{\dagger}_{\vec{x}+i})^2} 
\Big)
 \nn  \\
 +
 2&\text{Tr} \Big(
 \frac{h_1 W_{\vec{x}}}{(1 + h_1 W_{\vec{x}})^2} 
+  \frac{\bar{h}_1 W^{\dagger}_{\vec{x}}}{(1 + \bar{h}_1 W^{\dagger}_{\vec{x}})^2} 
\Big) 
\text{Tr}  \frac{\frac{1}{N_{\tau}-1} \sum_{t=1}^{N_{\tau}-1} (2 \kappa)^{2t}}{(1 + h_1 W_{\vec{x}+i}) (1 + \bar{h}_1 W^{\dagger}_{\vec{x}+i})}
\nn \\
+ 2 &\text{Tr} \frac{\frac{1}{N_{\tau}-1} \sum_{t=1}^{N_{\tau}-1} (2 \kappa)^{2t}}{(1 + h_1 W_{\vec{x}}) (1 + \bar{h}_1 W^{\dagger}_{\vec{x}})}
\text{Tr} \Big(
 \frac{h_1 W_{\vec{x}+i}}{(1 + h_1 W_{\vec{x}+i})^2} 
+ \frac{\bar{h}_1 W^{\dagger}_{\vec{x}+i}}{(1 + \bar{h}_1 W^{\dagger}_{\vec{x}+i})^2}  \Big) \nn \\
+ 2 \frac{1}{N_{\tau}-1} &\sum_{t=1}^{N_{\tau}-1} (2 \kappa)^{4t} \text{Tr}\frac{1}{(1 + h_1 W_{\vec{x}}) (1 + \bar{h}_1 W^{\dagger}_{\vec{x}})} \text{Tr}\frac{1}{(1 + h_1 W_{\vec{x}+i}) (1 + \bar{h}_1 W^{\dagger}_{\vec{x}+i})}
\nn \\
+ 2(2\kappa)^{2N_{\tau}} &\text{Tr}\frac{1}{(1 + h_1 W_{\vec{x}}) (1 + \bar{h}_1 W^{\dagger}_{\vec{x}})} \text{Tr}\frac{1}{(1 + h_1 W_{\vec{x}+i}) (1 + \bar{h}_1 W^{\dagger}_{\vec{x}+i})} \Big\}
\nn \\
 + 2 \frac{\kappa^4 N_{\tau}}{N_c^2-1} 
\sum_{\vec{x}, i} 
 \Big\{
\Big(
& \text{Tr} 
\frac{h_1 W_{\vec{x}}}{1 + h_1 W_{\vec{x}}} 
- \text{Tr}
\frac{\bar{h}_1 W^{\dagger}_{\vec{x}}}{1 + \bar{h}_1 W^{\dagger}_{\vec{x}}} 
\Big)^2 
\Big(
\text{Tr} 
\frac{h_1 W_{\vec{x}+i}}{1 + h_1 W_{\vec{x}+i}} 
- \text{Tr}
\frac{\bar{h}_1 W^{\dagger}_{\vec{x}+i}}{1 + \bar{h}_1 W^{\dagger}_{\vec{x}+i}} 
\Big)^2  \nn \\
&+ \text{Tr}
 \Big( \frac{h_1 W_{\vec{x}}}{1 + h_1 W_{\vec{x}}} 
- \frac{\bar{h}_1 W^{\dagger}_{\vec{x}}}{1 + \bar{h}_1 W^{\dagger}_{\vec{x}}} \Big)^2
\text{Tr}
 \Big( \frac{h_1 W_{\vec{x}+i}}{1 + h_1 W_{\vec{x}+i}} - \frac{\bar{h}_1 W^{\dagger}_{\vec{x}+i}}{1 + \bar{h}_1 W^{\dagger}_{\vec{x}+i}} \Big)^2
\Big\} \nn \\
 - 2 \frac{\kappa^4 N_{\tau}}{N_c^3-N_c} 
\sum_{\vec{x}, i} 
\Big\{
& \text{Tr}
 \Big( \frac{h_1 W_{\vec{x}}}{1 + h_1 W_{\vec{x}}} 
 - \frac{\bar{h}_1 W^{\dagger}_{\vec{x}}}{1 + \bar{h}_1 W^{\dagger}_{\vec{x}}} \Big)^2
\Big(
\text{Tr} 
\frac{h_1 W_{\vec{x}+i}}{1 + h_1 W_{\vec{x}+i}} 
- \text{Tr}
\frac{\bar{h}_1 W^{\dagger}_{\vec{x}+i}}{1 + \bar{h}_1 W^{\dagger}_{\vec{x}+i}} 
\Big)^2 \nn \\
 + 
\Big(
& \text{Tr} 
\frac{h_1 W_{\vec{x}}}{1 + h_1 W_{\vec{x}}} 
- \text{Tr}
\frac{\bar{h}_1 W^{\dagger}_{\vec{x}}}{1 + \bar{h}_1 W^{\dagger}_{\vec{x}}} 
\Big)^2 
\text{Tr}
 \Big( \frac{h_1 W_{\vec{x}+i}}{1 + h_1 W_{\vec{x}+i}}
 -\frac{\bar{h}_1 W^{\dagger}_{\vec{x}+i}}{1 + \bar{h}_1 W^{\dagger}_{\vec{x}+i}} \Big)^2 
\Big\} \nn
\end{align}

\begin{align}
-\int [dU_k] & \sum_{\vec{x}, i} \text{Tr} P_{\vec{x}, i} P_{\vec{x}, i} M_{\vec{x}, i} M_{\vec{x}, i} = \\
 2 \frac{\kappa^4 N_{\tau} (N_{\tau} -1)}{N_c^2} 
&\sum_{\vec{x}, i} 
\text{Tr}  \Big(
\frac{h_1 W_{\vec{x}-i}}{1 + h_1 W_{\vec{x}-i}} - \frac{\bar{h}_1 W^{\dagger}_{\vec{x}-i}}{1 + \bar{h}_1 W^{\dagger}_{\vec{x}-i}}
\Big) \nn \\
&\text{Tr} \Big(
 \frac{h_1 W_{\vec{x}}}{(1 + h_1 W_{\vec{x}})^2} +  \frac{\bar{h}_1 W^{\dagger}_{\vec{x}}}{(1 + \bar{h}_1 W^{\dagger}_{\vec{x}})^2} - 2  \frac{ \frac{1}{N_{\tau}-1} \sum_{t=1}^{N_{\tau}-1} (2 \kappa)^{2t}}{(1 + h_1 W_{\vec{x}}) (1 + \bar{h}_1 W^{\dagger}_{\vec{x}})}
\Big) \nn \\
&\text{Tr} \Big(
 \frac{h_1 W_{\vec{x}+i}}{1 + h_1 W_{\vec{x}+i}} - \frac{\bar{h}_1 W^{\dagger}_{\vec{x}+i}}{1 + \bar{h}_1 W^{\dagger}_{\vec{x}+i}}
\Big)\nn  \\
-2 \frac{\kappa^4 N_{\tau}}{N_c^2} 
\sum_{\vec{x}, i}  
\Big\{
&\text{Tr} \Big(
 \frac{h_1 W_{\vec{x}-i}}{1 + h_1 W_{\vec{x}-i}} - \frac{\bar{h}_1 W^{\dagger}_{\vec{x}-i}}{1 + \bar{h}_1 W^{\dagger}_{\vec{x}-i}}
\Big)
\text{Tr}
 \Big(1 - \frac{h_1 W_{\vec{x}}}{1 + h_1 W_{\vec{x}}}  + 1 - \frac{\bar{h}_1 W^{\dagger}_{\vec{x}}}{1 + \bar{h}_1 W^{\dagger}_{\vec{x}}} \Big)^2 \nn \\
&\text{Tr} \Big(
 \frac{h_1 W_{\vec{x}+i}}{1 + h_1 W_{\vec{x}+i}} - \frac{\bar{h}_1 W^{\dagger}_{\vec{x}+i}}{1 + \bar{h}_1 W^{\dagger}_{\vec{x}+i}}
\Big)
\Big\} \nn
\end{align}
\begin{align}
-\int [dU_k] &\sum_{\vec{x}, i \neq j} \text{Tr} P_{\vec{x}, i} M_{\vec{x}, j} P_{\vec{x}, j} M_{\vec{x}, i} = \\
&  2 \frac{\kappa^4 N_{\tau} (N_{\tau} -1)}{N_c^2} 
\sum_{\vec{x},i \neq j} 
\text{Tr} \Big(
 \frac{h_1 W_{\vec{x}-i}}{1 + h_1 W_{\vec{x}-i}} -\frac{\bar{h}_1 W^{\dagger}_{\vec{x}-i}}{1 + \bar{h}_1 W^{\dagger}_{\vec{x}-i}}
\Big) \nn \\
&\text{Tr} \Big(
 \frac{h_1 W_{\vec{x}}}{(1 + h_1 W_{\vec{x}})^2} + \frac{\bar{h}_1 W^{\dagger}_{\vec{x}}}{(1 + \bar{h}_1 W^{\dagger}_{\vec{x}})^2}
\Big)
\text{Tr} \Big(
\frac{h_1 W_{\vec{x}+j}}{1 + h_1 W_{\vec{x}+j}} - \frac{\bar{h}_1 W^{\dagger}_{\vec{x}+j}}{1 + \bar{h}_1 W^{\dagger}_{\vec{x}+j}}
\Big) \nn \\
-\frac{\kappa^4 N_{\tau}}{N_c^2} 
\sum_{\vec{x},i \neq j} 
\Big\{
&\text{Tr} \Big(
\frac{h_1 W_{\vec{x}-i}}{1 + h_1 W_{\vec{x}-i}} - \frac{\bar{h}_1 W^{\dagger}_{\vec{x}-i}}{1 + \bar{h}_1 W^{\dagger}_{\vec{x}-i}}
\Big) \nn \\
&\text{Tr} \Big[
\Big(1 - \frac{h_1 W_{\vec{x}}}{1 + h_1 W_{\vec{x}}} + 1 - \frac{\bar{h}_1 W^{\dagger}_{\vec{x}}}{1 + \bar{h}_1 W^{\dagger}_{\vec{x}}}\Big)^2
+ \Big( \frac{h_1 W_{\vec{x}}}{1 + h_1 W_{\vec{x}}} -  \frac{\bar{h}_1 W^{\dagger}_{\vec{x}}}{1 + \bar{h}_1 W^{\dagger}_{\vec{x}}} \Big)^2
\Big] \nn \\
&\text{Tr}  \Big(
\frac{h_1 W_{\vec{x}+j}}{1 + h_1 W_{\vec{x}+j}} - \frac{\bar{h}_1 W^{\dagger}_{\vec{x}+j}}{1 + \bar{h}_1 W^{\dagger}_{\vec{x}+j}}
\Big) 
\Big\}\nn
\end{align}
The $\text{Tr} P_i M_i P_j M_j$ and $\text{Tr} P_i M_j P_j M_i$ contributions are the same, just with different directions and a factor of $\frac12$ in front.
\begin{align}
\frac12 \int [dU_k] & \sum_{\vec{x}, i,j} \text{Tr} P_{\vec{x},i} M_{\vec{x},i} \text{Tr} P_{\vec{y},j} M_{\vec{y},j} = \\ 
& 2 \frac{\kappa^4 N_{\tau}^2}{N_c^2} \sum_{\vec{x}, i,j} 
 \Big( 
\text{Tr}\frac{h_1 W_{\vec{x}}}{1 + h_1 W_{\vec{x}}} 
- \text{Tr}\frac{\bar{h}_1 W^{\dagger}_{\vec{x}}}{1 + \bar{h}_1 W^{\dagger}_{\vec{x}}}  \Big)^2 
\nn \\
& \text{Tr} \Big(
\frac{h_1 W_{\vec{x}+i}}{1 + h_1 W_{\vec{x}+i}} 
- \frac{\bar{h}_1 W^{\dagger}_{\vec{x}+i}}{1 + \bar{h}_1 W^{\dagger}_{\vec{x}+i}}
\Big)
\text{Tr}\Big(
\frac{h_1 W_{\vec{x}+j}}{1 + h_1 W_{\vec{x}+j}} 
- \frac{\bar{h}_1 W^{\dagger}_{\vec{x}+j}}{1 + \bar{h}_1 W^{\dagger}_{\vec{x}+j}} 
\Big) \nn
\end{align}


\newpage
\begin{thebibliography}{99}

\bibitem{deForcrand:2010ys}
  P.~de Forcrand,
  PoS LAT {\bf 2009} (2009) 010
  [arXiv:1005.0539 [hep-lat]].
 
\bibitem{Aarts:2013bla}
  G.~Aarts,
  PoS LATTICE {\bf 2012} (2012) 017
  [arXiv:1302.3028 [hep-lat]].
  
\bibitem{Aarts:2013uxa}
  G.~Aarts, L.~Bongiovanni, E.~Seiler, D.~Sexty and I.~-O.~Stamatescu,
  Eur.\ Phys.\ J.\ A {\bf 49} (2013) 89
  [arXiv:1303.6425 [hep-lat]].
  
\bibitem{Gattringer:2012df}
  C.~Gattringer and T.~Kloiber,
  Nucl.\ Phys.\ B {\bf 869} (2013) 56
  [arXiv:1206.2954 [hep-lat]].
   
\bibitem{Delgado:2012tm}
  Y.~D.~Mercado, C.~Gattringer and A.~Schmidt,
  Comput.\ Phys.\ Commun.\  {\bf 184} (2013) 1535
  [arXiv:1211.3436 [hep-lat]].
  
\bibitem{Cristoforetti:2012su}
  M.~Cristoforetti {\it et al.}  [AuroraScience Collaboration],
  Phys.\ Rev.\ D {\bf 86} (2012) 074506
  [arXiv:1205.3996 [hep-lat]].

\bibitem{denes}
D.~Sexty,
  arXiv:1307.7748 [hep-lat].

\bibitem{Langelage:2010yr}
  J.~Langelage, S.~Lottini and O.~Philipsen,
  JHEP {\bf 1102} (2011) 057
   [Erratum-ibid.\  {\bf 1107} (2011) 014]
  [arXiv:1010.0951 [hep-lat]].

\bibitem{Fromm:2011qi}
  M.~Fromm, J.~Langelage, S.~Lottini and O.~Philipsen,
  JHEP {\bf 1201} (2012) 042
  [arXiv:1111.4953 [hep-lat]].
 
\bibitem{Fromm:2012eb}
  M.~Fromm, J.~Langelage, S.~Lottini, M.~Neuman and O.~Philipsen,
  Phys.\ Rev.\ Lett.\  {\bf 110} (2013) 122001
  [arXiv:1207.3005 [hep-lat]].

\bibitem{procs}
J.~Langelage, M.~Neuman and O.~Philipsen,
  arXiv:1311.4409 [hep-lat].
  
\bibitem{test}
G.~Bergner, J.~Langelage and O.~Philipsen,
  JHEP {\bf 1403} (2014) 039
  [arXiv:1312.7823 [hep-lat]].
        
 \bibitem{pinke}
 O.~Philipsen and C.~Pinke,
  arXiv:1402.0838 [hep-lat].
     
\bibitem{Unger:2011it}
  W.~Unger and P.~de Forcrand,
  J.\ Phys.\ G {\bf 38} (2011) 124190
  [arXiv:1107.1553 [hep-lat]].
  
\bibitem{Fromm:2011kq}
  M.~Fromm, J.~Langelage, O.~Philipsen, P.~de Forcrand, W.~Unger and K.~Miura,
  PoS LATTICE {\bf 2011} (2011) 212
  [arXiv:1111.4677 [hep-lat]].
 
\bibitem{Kawamoto:2005mq}
  N.~Kawamoto, K.~Miura, A.~Ohnishi and T.~Ohnuma,
  Phys.\ Rev.\ D {\bf 75} (2007) 014502
  [hep-lat/0512023].
  
\bibitem{Nakano:2010bg}
  T.~Z.~Nakano, K.~Miura and A.~Ohnishi,
  Phys.\ Rev.\ D {\bf 83} (2011) 016014
  [arXiv:1009.1518 [hep-lat]].
 
\bibitem{poly1}
C.~Wozar, T.~Kaestner, A.~Wipf and T.~Heinzl,
  Phys.\ Rev.\ D {\bf 76} (2007) 085004
  [arXiv:0704.2570 [hep-lat]].

\bibitem{poly2}
D.~Smith, A.~Dumitru, R.~Pisarski and L.~von Smekal,
  Phys.\ Rev.\ D {\bf 88} (2013) 054020
  [arXiv:1307.6339 [hep-lat]].
  
\bibitem{poly3}
J.~Greensite and K.~Langfeld,
  Phys.\ Rev.\ D {\bf 88} (2013) 074503
  [arXiv:1305.0048 [hep-lat]].

\bibitem{poly4}
J.~Greensite and K.~Langfeld,
  Phys.\ Rev.\ D {\bf 87} (2013) 094501
  [arXiv:1301.4977 [hep-lat]].

\bibitem{poly5}
C.~S.~Fischer, L.~Fister, J.~Luecker and J.~M.~Pawlowski,
  arXiv:1306.6022 [hep-ph].
 
\bibitem{Langelage:2010nj}
  J.~Langelage, S.~Lottini and O.~Philipsen,
  PoS LATTICE {\bf 2010} (2010) 196
  [arXiv:1011.0095 [hep-lat]].
  
\bibitem{Creutz:1978ub}
  M.~Creutz,
  J.\ Math.\ Phys.\  {\bf 19} (1978) 2043.

\bibitem{son}
D.~T.~Son and M.~A.~Stephanov,
  Phys.\ Rev.\ Lett.\  {\bf 86} (2001) 592
  [hep-ph/0005225].
  
 \bibitem{dh}
P.~H.~Damgaard and H.~H\"uffel,
  Phys.\ Rept.\  {\bf 152}, 227 (1987).

\bibitem{clsu3}
 F.~Karsch and H.~W.~Wyld,
  Phys.\ Rev.\ Lett.\ {\bf 55}, 2242 (1985)
  
   \bibitem{bilic88}
 N.~Bili\'{c},
  Phys.\ Rev.\ D {\bf 37}, 3684 (1988)
  
\bibitem{etiology}
G.~Aarts, F.~A.~James, E.~Seiler and I.~-O.~Stamatescu,
  Eur.\ Phys.\ J.\ C {\bf 71}, 1756 (2011)
  [arXiv:1101.3270 [hep-lat]].

\bibitem{su3lang}
G.~Aarts and F.~A.~James,
  JHEP {\bf 1201}, 118 (2012)
  [arXiv:1112.4655 [hep-lat]].
  
  \bibitem{gross83}
 M.~Gross, J.~Bartholomew and D.~Hochberg,
  Report No. EFI-83-35-CHICAGO, 1983
  
    \bibitem{adaptive-stepsize}
G.~Aarts, F.~A.~James, E.~Seiler and I.~-O.~Stamatescu,
  Phys. Lett. B 687 (2010)
   arXiv:0912.0617 [hep-lat].

  \bibitem{amb86}
J.~Ambjorn, M.~Flensburg and C.~Peterson,
  Nucl.\ Phys.\ B {\bf 275} (1986) 375.
  
  \bibitem{kim13}
 A.~Mollgaard and K.~Splittorff,
  Phys.\ Rev.\ D {\bf 88} (2013) 116007
  [arXiv:1309.4335 [hep-lat]].

\bibitem{cool}
 E.~Seiler, D.~Sexty and I.~-O.~Stamatescu,
  Phys.\ Lett.\ B {\bf 723} (2013) 213
  [arXiv:1211.3709 [hep-lat]].

\bibitem{hqet}
J.~Heitger {\it et al.}  [ALPHA Collaboration],
  JHEP {\bf 0402} (2004) 022
  [hep-lat/0310035].
    
\bibitem{sommer}  
S.~Necco and R.~Sommer,
  Nucl.\ Phys.\ B {\bf 622}, 328 (2002)
  [hep-lat/0108008].

\bibitem{cohen}
T.~D .Cohen,
  Phys.\ Rev.\ Lett.\  {\bf 91} (2003) 222001
  [hep-ph/0307089].

\end{thebibliography}
\end{document}